\documentclass[%
aip,
pof,
superscriptaddress
]{revtex4-1}

\usepackage{graphicx}
\usepackage{dcolumn}
\usepackage{epstopdf}
\usepackage{bm}
\usepackage[titletoc]{appendix}

\begin{document}






\title{A well-balanced gas kinetic scheme for Navier-Stokes equations with gravitational potential}
\author{Songze Chen}
\email{jacksongze@hust.edu.cn}
\affiliation{
State Key Laboratory of Coal Combustion, School of Energy and Power Engineering,
Huazhong University of Science and Technology, Wuhan, China
}%
\author{Zhaoli Guo}
\affiliation{
State Key Laboratory of Coal Combustion, School of Energy and Power Engineering,
Huazhong University of Science and Technology, Wuhan, China
}%
\author{Kun Xu}%
\affiliation{
The Hong Kong University of Science and technology,
Clear Water Bay, Kowloon, Hong Kong, China
}%

\date{\today}

\begin{abstract}
The hydrostatic equilibrium state is the consequence of the exact hydrostatic balance between hydrostatic pressure and external force.
Standard finite volume or finite difference schemes cannot keep this balance exactly due to their unbalanced truncation errors.
In this study, we introduce an auxiliary variable which becomes constant at isothermal hydrostatic equilibrium state and propose a well-balanced gas kinetic scheme for the Navier-Stokes equations with a global reconstruction.
Through reformulating the convection term and the force term via the auxiliary variable, zero numerical flux and zero numerical source term are enforced at the hydrostatic equilibrium state instead of the balance between hydrostatic pressure and external force.
Several problems are tested numerically to demonstrate the accuracy and the stability of the new scheme, and the results confirm that,
the new scheme can preserve the exact hydrostatic solution. The small perturbation riding on hydrostatic equilibria can be calculated accurately. The viscous effect is also illustrated through the propagation of small perturbation and the Rayleigh-Taylor instability.
More importantly, the new scheme is capable of simulating the process of converging towards hydrostatic equilibrium state from a highly non-balanced initial condition. The ultimate state of zero velocity and constant temperature is achieved up to machine accuracy.
As demonstrated by the numerical experiments, the current scheme is very suitable for small amplitude perturbation and long time running under gravitational potential.

\end{abstract}

\maketitle


\section{Introduction}
Gravity is involved in many physical problems, including astrophysical problems like core-collapse supernova, atmospheric motions on planet, smoke stratification in compartment fires etc.
In order to understand these phenomena and make reliable prediction, conservation laws with gravitational force are invoked in the form of partial differential equations.
However, numerical simulations of these systems are not easy from the following aspects: (1) for a long time evolution, the truncation error will accumulate and dramatically affect the final solution of an isolated gravitational system \cite{tian2007gravity,xing2013high}; (2) to predict small perturbation, say, numerical weather prediction and climate modeling, the truncation error will mask the small perturbations on the top of stationary solution \cite{xing2013high,chandrashekar2015second}.

These two problems can be attributed to unbalanced discretization of the convection term and gravitational force term.
Consider a fluid system under gravity governed by the Euler equations. The fluid system possesses a stationary state known as the hydrostatic equilibrium state in which the gravitational force is exactly balanced by the pressure gradient.
However, in conventional numerical schemes, the gravitational force term and the convection term are discretized separately, thereby, the truncation errors cannot cancel each other. As a result, the conventional numerical schemes are not able to preserve the hydrostatic equilibrium state, and can induce unacceptable spurious motions \cite{botta2004well}. In this context, a numerical scheme which ensures the hydrostatic balance exactly on discrete level is termed well-balanced.

Many well-balanced schemes has been developed, especially for shallow water equations \cite{audusse2000kinetic, leveque1998balancing, xing2005high}.
But the techniques developed for shallow water equations seem not easy to be implemented in the Euler equations with gravitational force term. This problem bothers the CFD community for a long time.

Botta et al.\cite{botta2004well} developed a well-balanced finite volume method for the Euler equations, using a discrete Archimedes
principle to express the gravity source term as the cell surface integral of the reconstructed
hydrostatic pressure.
With the help of kinetic theory, Xu et al.\cite{xu2010wbEuler} proposed a kind of well-balanced scheme for the Euler equations in 2010.
The gravitational potential is approximated as a step function inside each cell, and the amount of particle penetration and reflection from the cell interface is evaluated according to the incident particle velocity and the strength of the potential barrier at the cell interface.
This scheme can maintain hydrostatic equilibrium state exactly if the numerical integration used in kinetic flux is evaluated accurately.

In recent years, the development of well-balanced schemes for the Euler equations has gained more attention.
In 2012, Xing and Shu \cite{xing2013high} developed a special source term discretization so that the resulting WENO scheme balances the zero-velocity and constant temperature steady state solutions to machine accuracy, and at the same time maintains the high order accuracy and essentially non-oscillatory property for general solutions.
Unlike the kinetic scheme with step potential assumption, their scheme is free of analytical or numerical integration.
K\"{a}ppeli and Mishra \cite{kappeli2014wb} developed a novel reconstruction of the enthalpy which is based on local constant entropy assumption. After that they proposed a more general pressure reconstruction using a local analytical integration of hydrostatic equation, and demonstrated the efficiency of their well-balanced schemes for a broad set of astrophysical scenarios with several types of equation of state \cite{kappeli2016wb}.
Ghosh and Constantinescu \cite{ghosh2015wb,ghosh2016wb} extended Xing and Shu's work to more general flows encountered in atmospheric simulations.
Besides an isothermal equilibrium, the proposed well-balanced scheme can hold for many other hydrostatic equilibrium states.
Chandrashekar and Klingenberg \cite{chandrashekar2015second} also proposed a well-balanced scheme which involves a specific combination of source term discretization similar to Xing and Shu's, and requests that numerical flux exactly resolves stationary contacts.
The scheme is able to preserve isothermal and polytropic stationary solutions up to machine precision.
Chertock et al. \cite{chertock2018} proposed a global flux which reflects the accumulating effect of the gravitational potential along a direction and developed a well-balanced scheme for the Euler equations.

For the Navier-Stokes equations, the well-balanced scheme is rarely reported except the gas kinetic scheme(GKS) \cite{luo2011wb}.
As early as last century, Slyz and Prendergast \cite{slyz1999time} incorporated a time-independent gravitational potential into the gas kinetic scheme and guaranteed conservation of total (kinetic+internal+gravitational) energy, but did not consider the gravity in numerical flux.
Tian et al \cite{tian2007gravity} introduced the gravitational source term into numerical flux through the Chapman-Enskog expansion of the BGK equation.
Although many attempts have been made in the GKS framework, the GKS at that time were only of second-order accuracy with respect to holding the hydrostatic equilibrium state, and could not maintain the zero velocity and constant temperature to the machine zero.
In 2011, Luo et al\cite{luo2011wb} followed Xu's idea, used a piecewise constant function inside each cell to represent gravitational potential and the physical mechanism of particle transport across the potential barrier is explicitly used in the flux evaluation.
The proposed symplecticity-preserving gas kinetic scheme was proofed to be well-balanced for the Navier-Stokes equations (WB-NS, for short).
However, the disadvantages of the symplecticity-preserving gas kinetic scheme are also obvious: (1) the step piecewise representation of potential restricts the accuracy of numerical solution at a very low level; (2) potential barrier at cell interface complicates the computation of numerical flux.

The present study mainly focuses on the development of well-balanced gas kinetic scheme for the Navier-Stokes equations.
We are going to introduce an auxiliary variable which allows us to develop a simple and computational efficient well-balanced gas kinetic scheme under arbitrary potential function. Through the evolution towards the equilibrium state, we will show the difference between the equilibrium states of the Euler equations and the Navier-Stokes equations.

The remaining part is organized as follows: section 2 briefly introduces the auxiliary variables and modified the source terms in hydrodynamic equations; section 3 introduces the numerical flux calculation and the interpolation of the auxiliary variables; section 4 discusses the well-balanced property and the convergence towards isothermal hydrostatic equilibrium state through several numerical tests; section 5 concludes this study.

\section{Auxiliary variable and modified source terms}
\subsection{Two strategies to eliminate the truncation errors}
Truncation error is inevitable in many numerical schemes.
In order to eliminate the truncation error, Xing and Shu \cite{xing2013high} proposed a strategy in which the convection term and the source term are discretized by the same difference scheme. Although every single discretization generates truncation error, identical difference scheme guarantees the truncation error cancels each other. This strategy can be labeled as generating and eliminating.

On the contrary, another strategy is to prevent the occurrence of truncation error.
As we know, zero truncation error only occurs under special circumstance, say, the discretization of a constant function.
For example, consider the one dimensional Euler equations without external force,
\begin{eqnarray}
\left\{\begin{array}{rcc}
\rho_t + (\rho U)_x &=& 0, \\
(\rho U)_t + (\rho U^2 + p)_x &=& 0, \\
\displaystyle(\frac{1}{2}(\rho U^2 + \frac{2}{\gamma-1}p))_t + (U\frac{1}{2}(\rho U^2 + \frac{2\gamma}{\gamma-1}p))_x &=& 0,
\end{array}
\right.
\end{eqnarray}
where $\rho$ represents the density, $U$ represents the velocity, $p$ denotes the pressure, and $\gamma$ denotes the specific heat ratio.
The corresponding hydrostatic equilibrium state is trivial as the density, velocity and pressure are uniform everywhere.
\begin{equation}
\rho = \rho_{ref}, \quad U = U_{ref}, \quad p = p_{ref}
\end{equation}
In fact, every available numerical schemes are well-balanced for this hydrostatic equilibrium state.
\begin{figure}[phtb]
\centering
\includegraphics[width=0.8\textwidth]{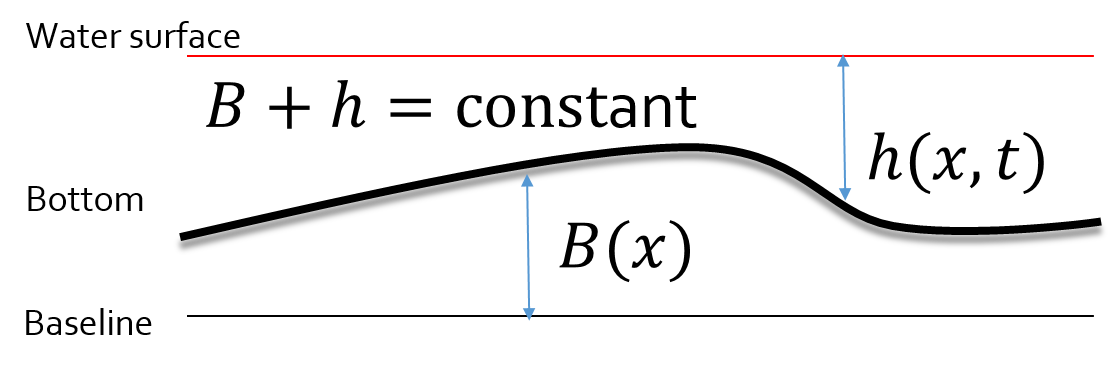}
\caption{Flat water surface at hydrostatic equilibrium state, where $h+B=\text{constant}$, }
\label{fig:shallowWater}
\end{figure}
Another example is the hydrostatic solution for the shallow water equations,
\begin{eqnarray}
\left\{
\begin{array}{rcc}
h_t+(hU)_x &=& 0, \\
(hU)_t+(hU^2 + \frac{1}{2}gh^2)_x &=& -ghB_x,
\end{array}
\right.
\end{eqnarray}
where $h$ represents the water height above this bottom, $U$ is the velocity, $g$ is the gravitational constant, and $B$ is the bottom elevation.
The top surface is at $B+h$ which is a constant at hydrostatic equilibrium.
By taking advantage of constant surface elevation ($B+h$), a kind of well-balanced scheme is developed \cite{xu2002shallowWater,zhou2001surfaceGradient}.

However, constant function will not happen naturally in most cases.
Consider the Euler equations under an external potential,
\begin{eqnarray}
\left\{
\begin{array}{rcc}
\rho_t + (\rho U)_x &=& 0, \\
(\rho U)_t + (\rho U^2 + p)_x &=& -\phi_x \rho \label{eq:EulerWith},\\
\displaystyle(\frac{1}{2}(\rho U^2 + \frac{2}{\gamma-1}p))_t + (\frac{1}{2}U(\rho U^2 + \frac{2\gamma}{\gamma-1}p))_x &=& -\phi_x (\rho U),
\end{array}
\right.
\end{eqnarray}
where $\phi$ is the external potential. 
If ideal gas state equation is adopted, $p=\rho RT$, where $R$ is the gas constant and $T$ is the temperature,
the isothermal hydrostatic equilibrium state is
\begin{eqnarray}
\rho = \rho_{ref} \exp(-\frac{\phi}{RT_{ref}}),\quad U = 0,\quad T = T_{ref}.
\label{eq:hydrostaticEq}
\end{eqnarray}
Since the exponential function cannot be accurately approximated by numerical discretizations based on Taylor expansion (polynomial), many numerical schemes cannot exactly hold this hydrostatic equilibrium state.

\subsection{Auxiliary variable}
Inspired by Zhou et al.'s work\cite{zhou2001surfaceGradient}, we adopt the second strategy to develop a well-balanced scheme for the Navier-Stokes equations with external force.
To do so, we first introduce an auxiliary variable, $\alpha$, which varies with location and satisfies the following equation,
\begin{eqnarray}
\rho = \rho_{ref} \exp(-\frac{\phi}{\alpha}). \label{eq:alphaDef}
\end{eqnarray}
where $\rho_{ref}$ is a constant reference density. $\alpha$ can be regarded as an analogue of $RT$ based on dimensional analysis.
More importantly, when the gas system rests at hydrostatic equilibrium state, the temperature is uniform throughout the space for isolated systems. Therefore, $\alpha$ is also a constant if $\rho_{ref}$ is chosen properly.

In numerical scheme, the interpolation of density can be replaced by the interpolations of $\alpha$ and potential function $\phi$ through Eq.(\ref{eq:alphaDef}).
Therefore, we can circumvent the interpolation of the exponential function, and only deal with a constant function at the hydrostatic equilibrium state.

\subsection{Modified source terms}
The stationary Euler equation with external potential is written as follows,
\begin{eqnarray}
\left\{\begin{array}{rcl}
\displaystyle(\rho U)_x &=& 0, \\
\displaystyle(\rho U^2 + p)_x &=& -\phi_x \rho \label{eq:staticEuler},\\
\displaystyle(\frac{1}{2}U(\rho U^2 + \frac{2\gamma}{\gamma-1}p))_x &=& -\phi_x (\rho U).
\end{array}\right.
\end{eqnarray}
It is obvious that the mass and energy equations will be satisfied at hydrostatic equilibrium state because of zero velocity.
But the flux of momentum is nonzero and balanced by the external force.
Actually, the momentum equation is the only obstacle for numerically holding the hydrostatic equilibrium state.

If the external force term are discretized directly in the momentum equation,
the nonlinear distributed density and pressure will introduce nonzero truncation error and then induce nonzero velocity or oscillate the solution.
In order to remove the truncation error,
we propose a new strategy to calculate the source term.
Substitute Eq.(\ref{eq:alphaDef}) into the source term in the momentum equation,
\begin{eqnarray}
\rho\phi_x &=& (-\rho\alpha)_x + \rho\alpha_x(1+\frac{\phi}{\alpha}). \label{eq:transform}
\end{eqnarray}
Then reformulate the steady momentum equation as follows,
\begin{eqnarray}
(\rho U^2+p-\rho\alpha)_x + \rho\alpha_x(1+\frac{\phi}{\alpha}) = 0.   \label{eq:modifiedMomEq}
\end{eqnarray}
At the hydrostatic equilibrium state, the velocity is zero, thereby the Eq.(\ref{eq:modifiedMomEq}) becomes,
\begin{eqnarray}
(p-\rho\alpha)_x + \rho\alpha_x(1+\frac{\phi}{\alpha}) = 0.   \label{eq:steadyMomEq}
\end{eqnarray}
The first term is solely composed of the static pressure and $-\rho\alpha $ which can be taken as a negative pressure ($-p$) and will completely cancel the static pressure. As a result, the first term on left hand side of the Eq.(\ref{eq:steadyMomEq}) is zero.
The consequence is that the discretization of the first term will not introduce any truncation error at the hydrostatic equilibrium.
The last term on the left hand side of the Eq.(\ref{eq:steadyMomEq}) also vanishes since $\alpha$ is a constant.
Therefore the hydrostatic equilibrium state is exactly held by the above equation,
as long as the discretization can exactly approximate the zero derivative of the constant value of $\alpha$,
which can be easily fulfilled by most of numerical schemes.
One thing we should emphasize is that, the modified momentum equation will lead to non-conservative scheme, which is discussed in Appendix~\ref{app:non-conservative}.
In the following section, we will use the auxiliary variable and modified momentum equation (Eqs.(\ref{eq:alphaDef},\ref{eq:transform})) to construct a well-balanced scheme for
the Navier-Stokes equations.

\section{Well-balanced gas kinetic scheme with external force}
\label{sec:WBGKS}
\subsection{BGK equation with external force}
Consider the dimensionless BGK equation under external potential force in one dimensional space,
\begin{equation}
    \frac{\partial f}{\partial t}+u\frac{\partial f}{\partial x} - \phi_x\frac{\partial f}{\partial u} =
    \frac{g-f}{\tau},
    \label{eq:BGKModel}
\end{equation}
where $\tau$ denotes the relaxation time, $u$ denotes the particle velocity,
$f=f(x,u,\xi,t)$ represents the velocity distribution function, $g=g(x,u,\xi,t)$ represents the corresponding equilibrium state, the Maxwellian distribution function, which can be expressed as follows,
\begin{eqnarray}
g &=& \mathcal{M}(\bm{W}) = \rho\left(\frac{1}{2\pi RT}\right)^{(1+k)/2} e^{-\frac{(u-U)^2+\xi^2}{2RT}}. \label{eq:equilibrium}
\end{eqnarray}
where $\xi=(\xi_1,\xi_2,...,\xi_k)$ denotes the effective internal freedom and $k$ is the degree of effective internal freedom ($\gamma = \frac{k+3}{k+1}$ in one dimensional problem). The macroscopic variables can be derived by taking the moments of the microscopic distribution function,
\begin{eqnarray}
\bm{W}=\langle \bm{\psi} f \rangle &=&
\left(\begin{array} {c}
\displaystyle \int_{-\infty}^{+\infty}\int_{-\infty}^{+\infty} f dud\xi \\
\displaystyle \int_{-\infty}^{+\infty}\int_{-\infty}^{+\infty} u f dud\xi \\
\displaystyle \int_{-\infty}^{+\infty}\int_{-\infty}^{+\infty} \frac{1}{2}u^2 f dud\xi
\end{array}\right)
=\left(\begin{array} {c}
\rho \\
\rho U \\
\rho E
\end{array}\right),\quad
\bm{\psi} = \left(\begin{array} {c}
1 \\ u \\ \frac{1}{2}(u^2 + \xi^2)
\end{array}\right),
\end{eqnarray}
where $\rho E = \frac{1}{2}\rho(U^2+(k+1)RT)$ denotes the total energy.
The symbol $\langle f \rangle$ is defined as,
\begin{eqnarray}
\langle f \rangle = \int_{-\infty}^{+\infty}\int_{-\infty}^{+\infty} f d{u}d\xi.
\end{eqnarray}
The moments of $f_u$ can be expressed by lower order moments of $f$,
\begin{eqnarray}
\langle u^n\xi^m f_u \rangle &=& \int_{-\infty}^{+\infty}\int_{-\infty}^{+\infty} u^n\xi^m f_u dud\xi \nonumber\\
&=& -n\int_{-\infty}^{+\infty}\int_{-\infty}^{+\infty} u^{n-1}\xi^m f dud\xi = -n\langle u^{n-1}\xi^m f\rangle \label{eq:momOffu}
\end{eqnarray}
The conservation of collision term requires,
\begin{eqnarray}
\langle \bm{\psi} (g-f) \rangle = 0. \label{eq:compatibility}
\end{eqnarray}
Taking moments of the BGK equation, we have,
\begin{eqnarray}
\langle\psi f\rangle_t + \langle u\psi f\rangle_x - \phi_x \langle\psi f_u\rangle = 0 \quad \Rightarrow  \left\{\begin{array} {rcl}
\rho_t + (\mathcal{F}^\rho)_x &=& 0 \\
(\rho U)_t + (\mathcal{F}^{\rho U})_x +\phi_x \rho &=& 0 \\
(\rho E)_t + (\mathcal{F}^{\rho E})_x +\phi_x (\rho U) &=& 0 \end{array}\right.
\end{eqnarray}
Using the auxiliary variable to reformulate the above equations, we have,
\begin{eqnarray}
\left\{\begin{array} {rcl}
\displaystyle \rho_t + (\mathcal{F}^\rho)_x - \mathcal{S}^{\rho} &=& 0, \\
\displaystyle (\rho U)_t + (\mathcal{F}^{\rho U}-\rho\alpha)_x - \mathcal{S}^{\rho U} &=& 0,\\
\displaystyle (\rho E)_t + (\mathcal{F}^{\rho E})_x - \mathcal{S}^{\rho E} &=& 0. \end{array}\right.
\end{eqnarray}
The flux and the source term can be explicitly expressed as follows,
\begin{eqnarray}
\left.\begin{array} {lll}
\displaystyle \mathcal{F}^{\rho} = \langle uf \rangle, & \mathcal{F}^{\rho U} = \langle uuf \rangle, &
\mathcal{F}^{\rho E} = \langle u\frac{1}{2}(u^2+\xi^2)f \rangle, \\
\displaystyle \mathcal{S}^{\rho} = 0, & \mathcal{S}^{\rho U} = -\rho\alpha_x(1+\frac{\phi}{\alpha}),
& \mathcal{S}^{\rho E} = -\phi_x(\rho U)\end{array}\right.
\end{eqnarray}
In this study the Chapman-Enskog expansion is adopted in order to solve the Navier-Stokes equations,
$$f = g - \tau(g_t+ug_x-\phi_xg_u)+ O(\tau^2). $$
The second term on the right hand side corresponds to the Navier-Stokes constitutive relationship \cite{Chapman1970}.

\subsection{GKS flux with external force}
Consider the conservation law of $\bf{W}$ in a one dimensional control volume $\Delta x$ during time interval $\Delta t$,
\begin{eqnarray}
\bm{W}^{n+1}_i = \bm{W}^{n}_i - \frac{1}{\Delta x}(F_{i+1/2}-F_{i-1/2})+S_i, \quad i = 1,2,...,n \label{eq:fdm}
\end{eqnarray}
where $F$ denotes the numerical fluxes and $S$ represents the numerical source term during a time step.
As the modified equations present, the flux term and source term in this study become,
\begin{eqnarray}
F=\int_0^{\Delta t} \left(\begin{array} {c}
\mathcal{F}^{\rho} \\
\mathcal{F}^{\rho U} -\rho\alpha\\
\mathcal{F}^{\rho E}
\end{array}\right)dt =
\int_0^{\Delta t} \left(\langle u\bm{\psi} f \rangle  + \left(\begin{array} {c}
0 \\
\displaystyle -\rho_0\alpha_0\\
\displaystyle 0
\end{array}\right) \right)  dt , \label{eq:numericalF}
\end{eqnarray}
\begin{eqnarray}
S=\frac{1}{\Delta x}\int_0^{\Delta t}\int_{x_{i-\frac{1}{2}}}^{x_{i+\frac{1}{2}}}\left(\begin{array} {c}
\mathcal{S}^{\rho} \\
\mathcal{S}^{\rho U} \\
\mathcal{S}^{\rho E}
\end{array}\right)dxdt =
\frac{-1}{\Delta x}\int_0^{\Delta t} \int_{x_{i-\frac{1}{2}}}^{x_{i+\frac{1}{2}}}\left(\begin{array} {c}
0 \\
\displaystyle \rho_0\alpha_x(1+\frac{\phi}{\alpha_0})\\
\displaystyle -\phi_x(\rho U)
\end{array}\right)dxdt. \label{eq:numericalS}
\end{eqnarray}
The local approximate solution at cell interface is,
\begin{equation}
f(t) = g_0 - \tau(g_t+ug_x-\phi_x g_u) + g_t t,\label{eq:localSolution}
\end{equation}
where $g_0$ defined at a cell interface is the equilibrium state at the beginning of the time step.
The spatial derivative of the equilibrium state is calculated from the derivative of the macroscopic variables (Appendix~\ref{app:gks-formula}).
The moments involved with $g_u$ can be explicitly calculated by Eq.(\ref{eq:momOffu}).
Because of the conservation of the collision term, we have the following equation,
\begin{eqnarray}
\langle \psi (f-g) \rangle = \tau\langle\psi(g_t+ug_x-\phi_x g_u)\rangle = 0.
\end{eqnarray}
The time derivative of conservative variables $\langle \psi g_t \rangle$ can be derived,
\begin{eqnarray}
\langle\psi g_t \rangle = -\langle\psi (ug_x-\phi_x g_u) \rangle, \label{eq:gt}
\end{eqnarray}
Assume that $g_x$ and $\phi_x$ are constant during the time step
and conservative variables can be expressed as the Taylor expansion in terms of time.
Integrate Eq.(\ref{eq:gt}) over the time step,
and only retain leading order terms up to $O(\Delta t)$,
\begin{eqnarray}
\Delta \bm{W}^* = \left(\begin{array} {c}
\Delta\rho \\
\Delta(\rho U) \\
\Delta(\rho E)
\end{array}\right)
= - \Delta t\left(\begin{array} {c}
\langle ug_x \rangle \\
\langle uug_x \rangle +\phi_x \rho_0 \\
\langle\frac{1}{2} u(u^2+\xi^2) g_x\rangle + \phi_x \rho_0 U_0
\end{array}\right).
\end{eqnarray}
where subscript "0" denotes initial time at the beginning of the time step.
Then an intermediate equilibrium state at the end of the time step can be constructed,
\begin{eqnarray}
g^* = \mathcal{M}(\bm{W}^*) = \mathcal{M}(\bm{W}_0+\Delta \bm{W}^*).
\end{eqnarray}
This intermediate state is not the new state for the next time step,
but only used to estimate the numerical time derivative,
\begin{eqnarray}
g_t = \frac{g^*-g_0}{\Delta t}.
\end{eqnarray}
In fact, this procedure can be seen as the use of the Euler equations to predict the time derivative.
As all the terms in Eq.(\ref{eq:localSolution}) are derived, the numerical fluxes (Eq.(\ref{eq:numericalF}))
and the numerical source term (Eq.(\ref{eq:numericalS})) can be calculated.

\subsection{Discretization}
Suppose the computational domain is uniformly discretized by $n$ cells, and the cell size is $\Delta x$. The variables defined at cell $i$ are denoted by their subscript $i$ and the variables defined at the cell interface between $i$ and $i+1$ cells are denoted by subscript $i+1/2$.
\subsubsection{Reference density}
Reformulate Eq.(\ref{eq:alphaDef}) as follows,
\begin{eqnarray}
\alpha = \frac{\phi}{\ln\rho_{ref}-\ln\rho}. \label{eq:inverse}
\end{eqnarray}
Considering the above equation and Eq. (\ref{eq:alphaDef}), it requires that $\rho_{ref} \neq \rho$ and $\phi \neq 0$. Therefore, we define the reference density as follows,
\begin{eqnarray}
\ln\rho_{ref} = \ln\rho_{\max(\rho)} + \phi_{\max(\rho)} RT_{\max(\rho)},
\end{eqnarray}
where $\max(\rho)$ denotes the cell index whose density is maximum in the computational domain.
Furthermore, a positive value is added to the potential function to ensure its positivity throughout the entire computational domain.
$\rho_{ref}$ is calculated at the beginning of every time step globally.

\subsubsection{Auxiliary variables}
Instead of using the original variables in Eq. (\ref{eq:hydrostaticEq}),
\begin{eqnarray}
\bm{v}_i = \{\rho_i, \quad U_i, \quad T_i, \quad \phi_i\},   \label{eq:originalSet}
\end{eqnarray}
we perform interpolation with the following set of variables in which $\alpha$ instead of $\rho$ will be constant at the hydrostatic equilibrium state,
\begin{eqnarray}
\bar{\bm{v}}_i = \{\alpha_i, \quad U_i, \quad T_i, \quad \phi_i\}.  \label{eq:auxiliarySet}
\end{eqnarray}

Linear interpolation and central difference are adopted to approximate the interfacial values and corresponding derivatives respectively, that is,
\begin{eqnarray}
\bar{\bm{v}}_{i+1/2} &=& \frac{1}{2}(\bar{\bm{v}}_{i+1}+\bar{\bm{v}}_{i}), \\
\left.\frac{\partial \bar{\bm{v}}}{\partial x}\right|_{i+1/2} &=& \frac{\bar{\bm{v}}_{i+1}-\bar{\bm{v}}_{i}}{\Delta x}.
\end{eqnarray}
Then conservative variables and their the derivatives are derived from $\bar{\bm{v}}$ and $\frac{\partial \bar{\bm{v}}}{\partial x}$ by the transformation and the chain role in calculus. The quantities at cell interface (denoted by subscript "$i+1/2$") can be calculated via the interpolated values and derivatives.

The spatial integral of the source term is approximated by the trapezoidal rule of two interfacial values,
\begin{eqnarray}
S_i = -\frac{1}{2} \Delta t (\mathcal{S}_{i-1/2} + \mathcal{S}_{i+1/2}). \label{eq:sourceTermTrapezoid}
\end{eqnarray}
This procedure guarantees identical discretization is adopted for the flux and the source term, which is crucial for WB scheme as mentioned by Xing and Shu \cite{xing2013high}.

\subsection{Well-Balanced property}
Under isothermal hydrostatic equilibrium state ($\alpha = \alpha_{iso} = RT_{iso}, U=U_{iso}=0, T=T_{iso}, \phi$), the discretization introduced in last subsection will exactly reproduce the isothermal quantities, $\alpha_{iso}, U_{iso}, T_{iso}$ and their zero derivatives.  It is easy to verify the following identities (\textbf{Appendix A}),
\begin{eqnarray}
\langle \psi (ug_x-\phi_xg_u) \rangle = 0, \quad
\langle u\psi (ug_x-\phi_xg_u) \rangle = 0.
\end{eqnarray}
As a result, the time derivative, $g_t$ (Eq.(\ref{eq:gt})) will completely vanish at the hydrostatic equilibrium state.
Hence, the numerical fluxes (Eq.(\ref{eq:numericalF})) become,
\begin{eqnarray}
F_{i+1/2}&=&\int_0^{\Delta t} \left(\langle u\bm{\psi} (g_0-\tau(ug_x-\phi_xg_u)) \rangle  + \left(\begin{array} {c}
0 \\
\displaystyle -\rho_{i+1/2}\alpha_{iso}\\
\displaystyle 0
\end{array}\right) \right)  dt \nonumber\\
&=& \Delta t\left(\begin{array} {c}
0 \\
\displaystyle \rho_{i+1/2} (RT_{iso} - \alpha_{iso})\\
\displaystyle 0
\end{array}\right)
 = 0,
\end{eqnarray}
and the interfacial value of source term is zero too,
\begin{eqnarray}
\mathcal{S}_{i+1/2} = 0,
\end{eqnarray}
so is the volume integral (Eq.(\ref{eq:sourceTermTrapezoid}), $S_i = 0$).
The numerical source term and numerical flux term are both zero exactly. Therefore, the isothermal hydrostatic equilibrium state can be exactly held by the proposed numerical scheme regardless of the shape of $\phi$.


\section{Numerical results}
\subsection{Adiabatic Boundary condition}
The adiabatic boundary condition is adopted in order to exclude the effects of external heating when simulating the converging process towards the hydrostatic equilibrium state.
Under this circumstance, the entire computational domain becomes an isolated system.
We use ghost cell technique to realize adiabatic boundary condition.
The density, temperature and potential functions in the ghost cell at the boundary are assigned the same as those of the direct neighboring cell in the flow field,
and the fluid velocity is set to be the opposite of that in the neighboring cell in the flow field,
\begin{eqnarray}
\rho_g = \rho_f, \quad U_g = -U_f, \quad T_g = T_f, \quad \phi_g = \phi_f,
\end{eqnarray}
where subscript "$g$" represents the ghost cell, and "$f$" represents the flow field.
Thus no mass and no heat penetrate the solid boundaries. In the following numerical simulations, all the boundary conditions are adiabatic boundary condition, specific heat ratio of gas is 1.4 and gas constant is 1.0 if not specified. In most cases, the sound speed of the initial condition is $c_s=\sqrt{1.4}$; the length of computational domain is $L=1$; so the sound crossing time defined as $\tau = 2L/c_s$ is about $1.69$.

\subsection{Maintaining the isothermal hydrostatic equilibrium state}
\begin{figure}
\centering
    \parbox[b]{0.48\textwidth}{
    \includegraphics[totalheight=5.5cm, bb = 0 25 690 565, clip =
    true]{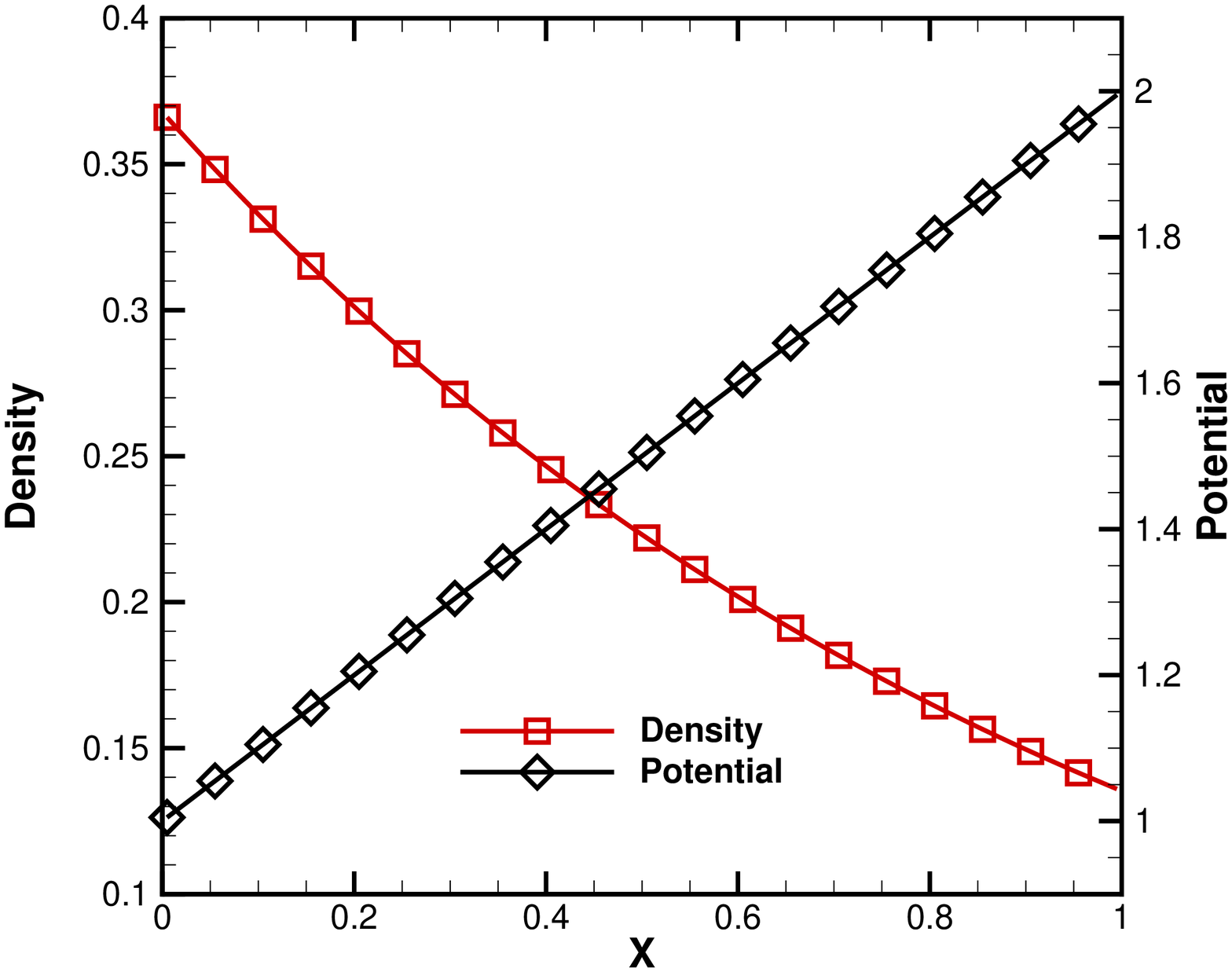}
    }
    \parbox[b]{0.48\textwidth}{
    \includegraphics[totalheight=5.5cm, bb = 0 25 690 565, clip =
    true]{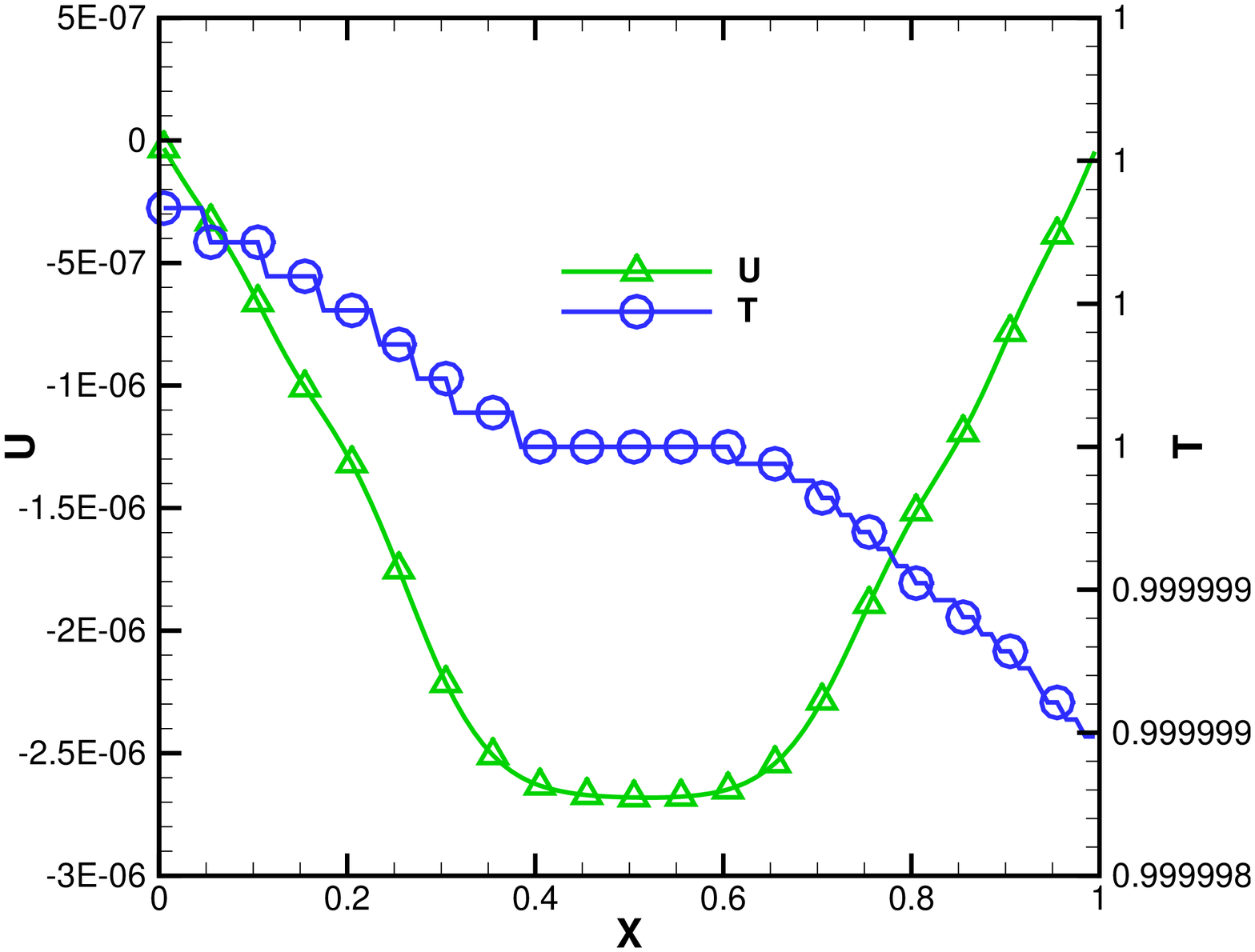}
    }
    \caption{The numerical results at $t=2.0$ for maintaining the isothermal hydrostatic equilibrium from non-well-balanced scheme.}
    \label{fig:nwbHoldEq}
\end{figure}
The well-balanced property of the present scheme is demonstrated firstly. We choose three different potentials to verify our code, and stop the simulations at time $t = 2.0$ to check whether the hydrostatic equilibrium state is kept.
The potential functions are given as follows,
\begin{eqnarray}
\phi_1(x) = x+1, \quad \phi_2(x) = x^2+1, \quad \phi_3(x) = \sin(2\pi x) + 2.   \label{eq:3-potentials}
\end{eqnarray}
The initial condition is given by Eq.(\ref{eq:hydrostaticEq}) with $\rho_{ref} = 1$ and $T_{ref} = 1$.
The computational domain is uniformly divided into 100 cells. Adiabatic boundary condition is adopted at the two ends of computational domain.

As a comparison, a primary scheme is employed for the simulation of $\phi_1(x) = x+1$ case, in which the conservative variables ($\rho, \rho U, \rho E$) and original source term are adopted, and the spatial integral of the Eq.(\ref{eq:numericalS}) is also calculated by trapezoidal rule. This scheme is referred as "nWB" hereafter. As shown in Fig. \ref{fig:nwbHoldEq}, the velocity cannot stay at zero and oscillates at the boundaries, and the temperature also deviates from the equilibrium condition, which implies the scheme is not well-balanced.

\begin{figure}[htb]
\centering
    \parbox[b]{0.48\textwidth}{
    \includegraphics[totalheight=5.5cm, bb = 0 25 690 565, clip =
    true]{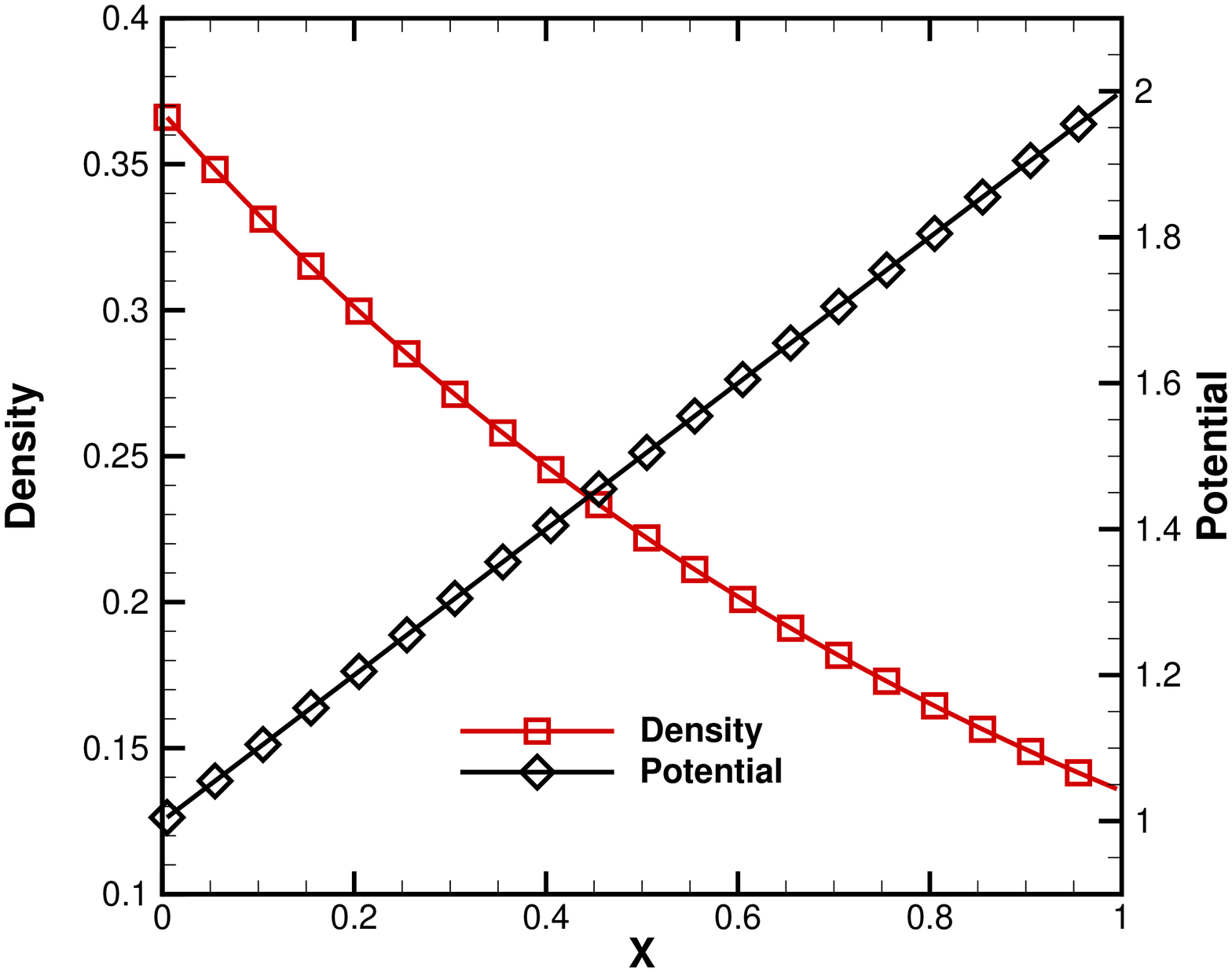}
    }
    \parbox[b]{0.48\textwidth}{
    \includegraphics[totalheight=5.5cm, bb = 0 25 690 565, clip =
    true]{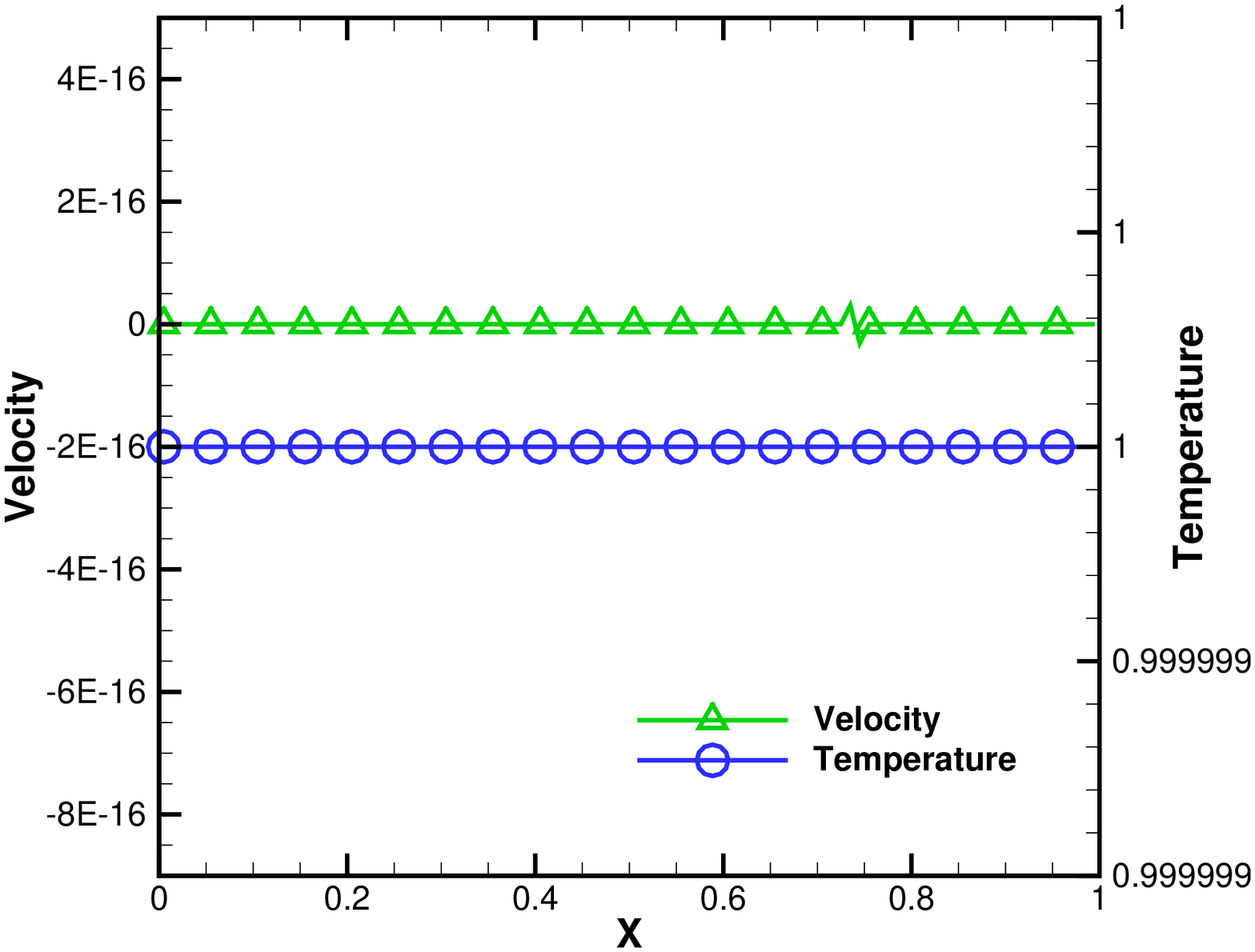}
    }
    \parbox[b]{0.48\textwidth}{
    \includegraphics[totalheight=5.5cm, bb = 0 25 690 565, clip =
    true]{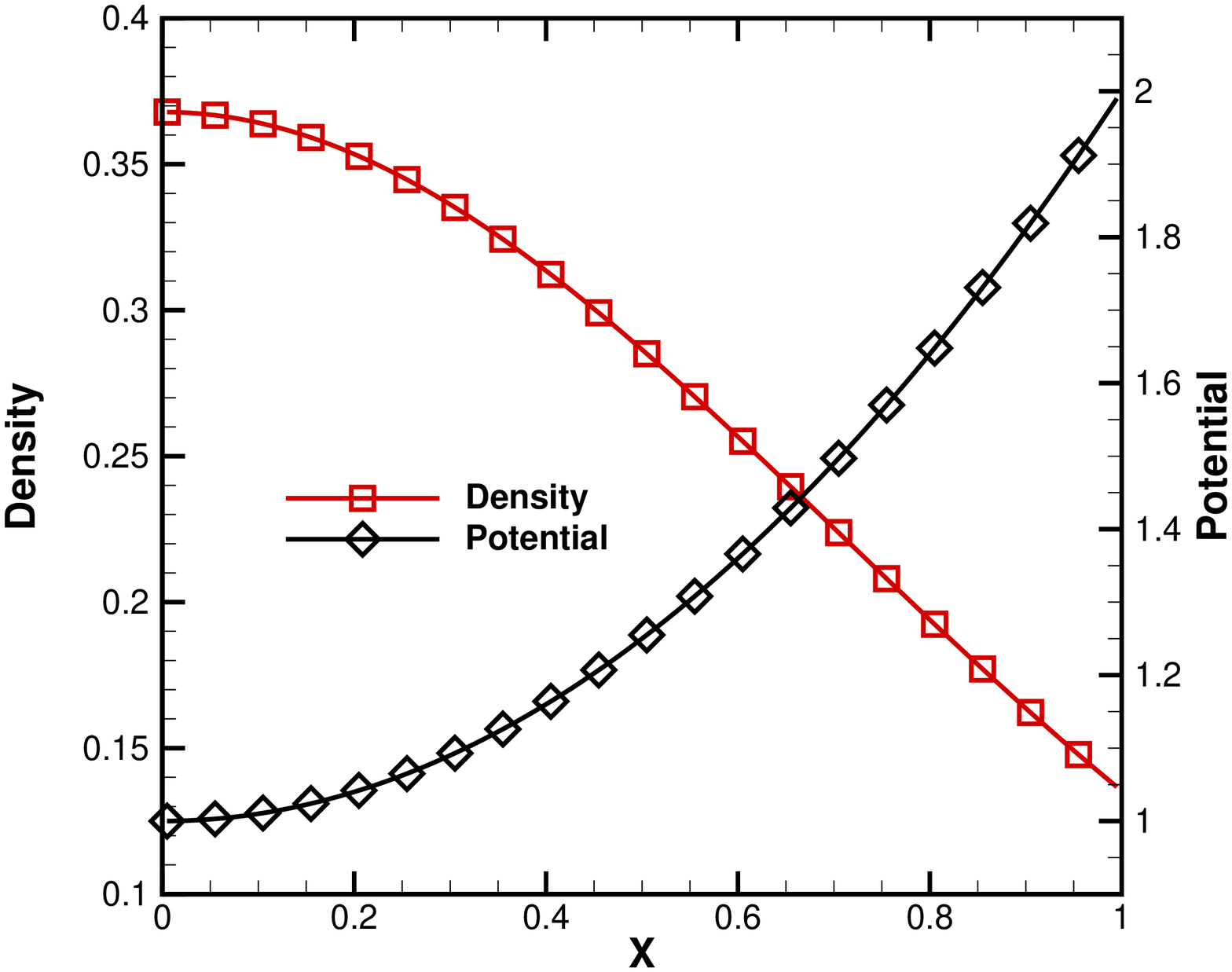}
    }
    \parbox[b]{0.48\textwidth}{
    \includegraphics[totalheight=5.5cm, bb = 0 25 690 565, clip =
    true]{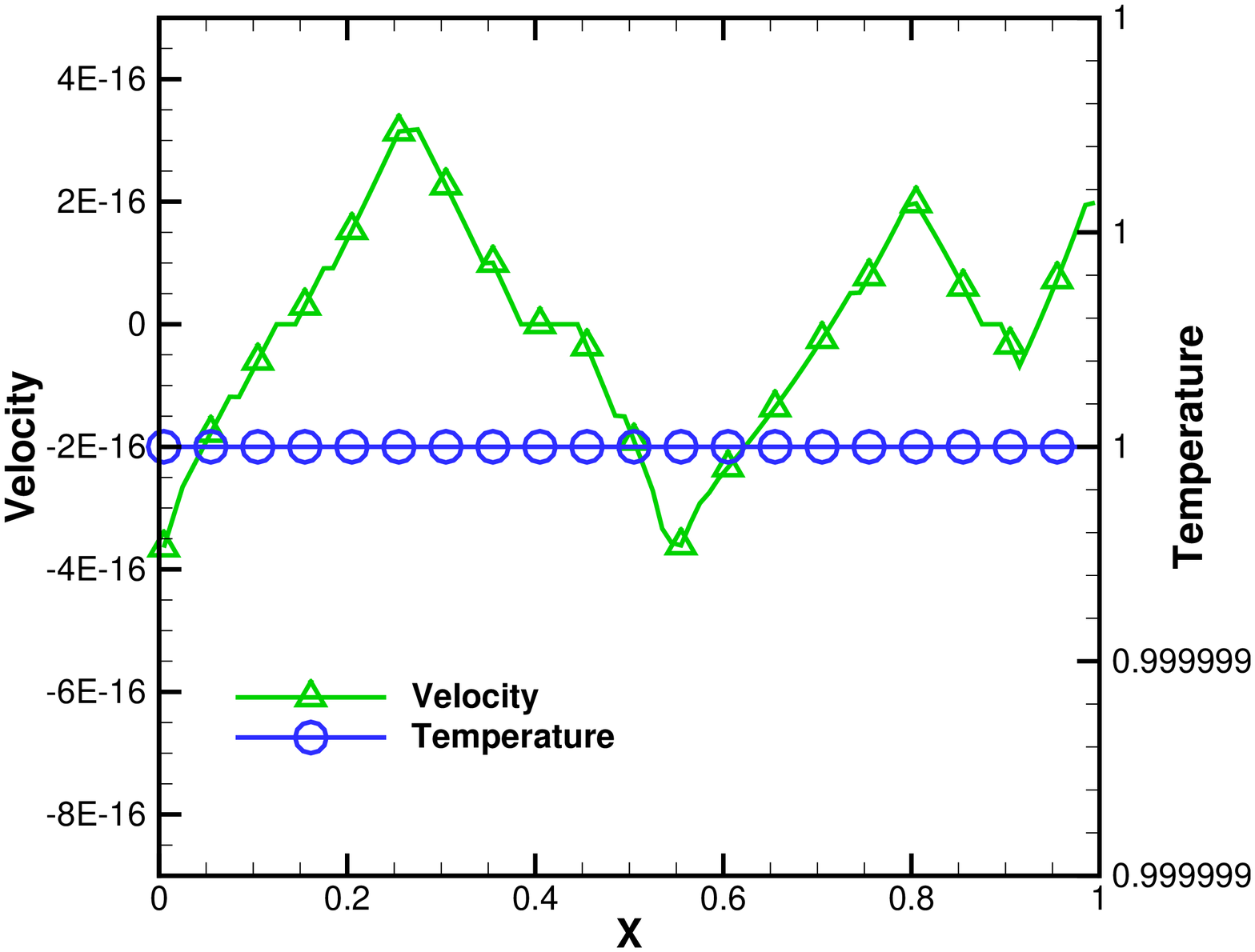}
    }
    \parbox[b]{0.48\textwidth}{
    \includegraphics[totalheight=5.5cm, bb = 0 25 690 565, clip =
    true]{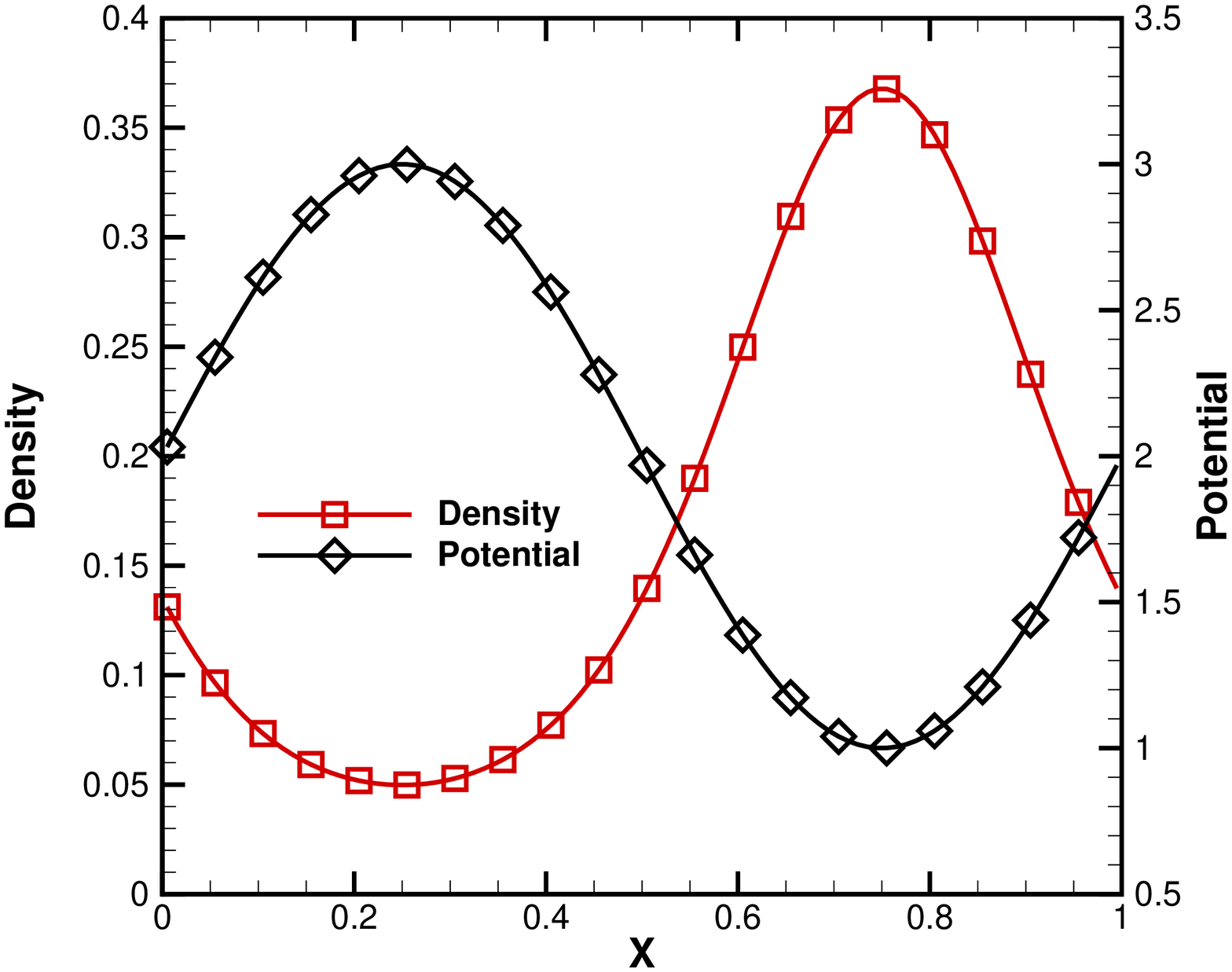}
    }
    \parbox[b]{0.48\textwidth}{
    \includegraphics[totalheight=5.5cm, bb = 0 25 690 565, clip =
    true]{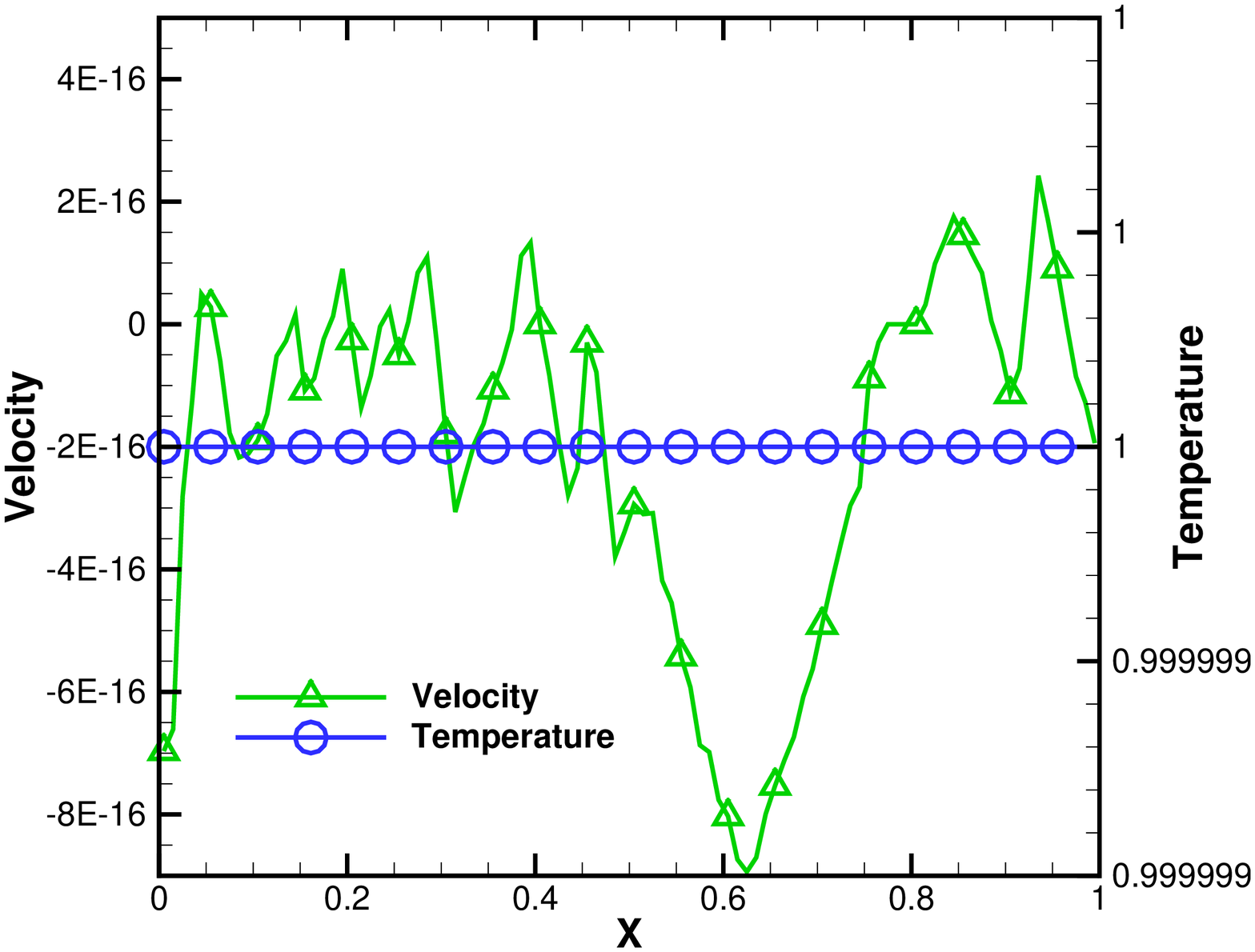}
    }
    \caption{The numerical results at $t=2.0$ for maintaining the isothermal hydrostatic equilibrium from present well-balanced scheme.
    The first row: $\phi_1(x) = x+1$; the second row: $\phi_2(x) = x^2+1$; the last row: $\phi_3(x) = \sin(2\pi x) + 2$}
    \label{fig:wbHoldEq}
\end{figure}

Then, auxiliary variables $\bar{\bm{v}}_i$ are used for the interpolation and the Eq.(\ref{eq:sourceTermTrapezoid}) is adopted for the spatial integral of the source term. As we expected, the modified equations and the new interpolation technique work perfectly. In Fig. \ref{fig:wbHoldEq} the hydrostatic equilibrium states are maintained up to the machine accuracy for all the potential functions, which verifies the well-balanced property.

\subsection{Propagation of perturbation on an isothermal hydrostatic equilibrium state}
Well-balanced property is of great importance for simulating the propagation of small perturbation riding on an hydrostatic equilibrium state. We simulate a test case from the reference\cite{leveque1999wave} to demonstrate the efficiency of the present scheme. Consider an ideal gas with $\gamma = 1.4$ staying initially at an isothermal hydrostatic equilibrium state,
$$ \rho_0(x) = p_0(x) = e^{-x}\quad \mbox{and}\quad U_0(x) = 0, $$
for $x \in [0,1]$. Then the initial pressure is perturbed by
\begin{eqnarray}
p(x, t=0) = p_0(x) + \eta \exp(-100(0.5-x)^2),
\end{eqnarray}
where $\eta$ is the amplitude of the perturbation.
The potential is given as follows,
\begin{eqnarray}
\phi(x) = x+1.
\end{eqnarray}
The computation is conducted with $100$ uniform cells in the whole domain and stops at time $t= 0.25$.
With $0 < \eta \ll 1$, the initial perturbation splits into two waves spreading on both sides.
\subsubsection{Inviscid flow}
\begin{figure}[htb]
\centering
    \parbox[b]{0.31\textwidth}{
    \centering (a)
    \includegraphics[totalheight=4.2cm, bb = 0 25 690 565, clip =
    true]{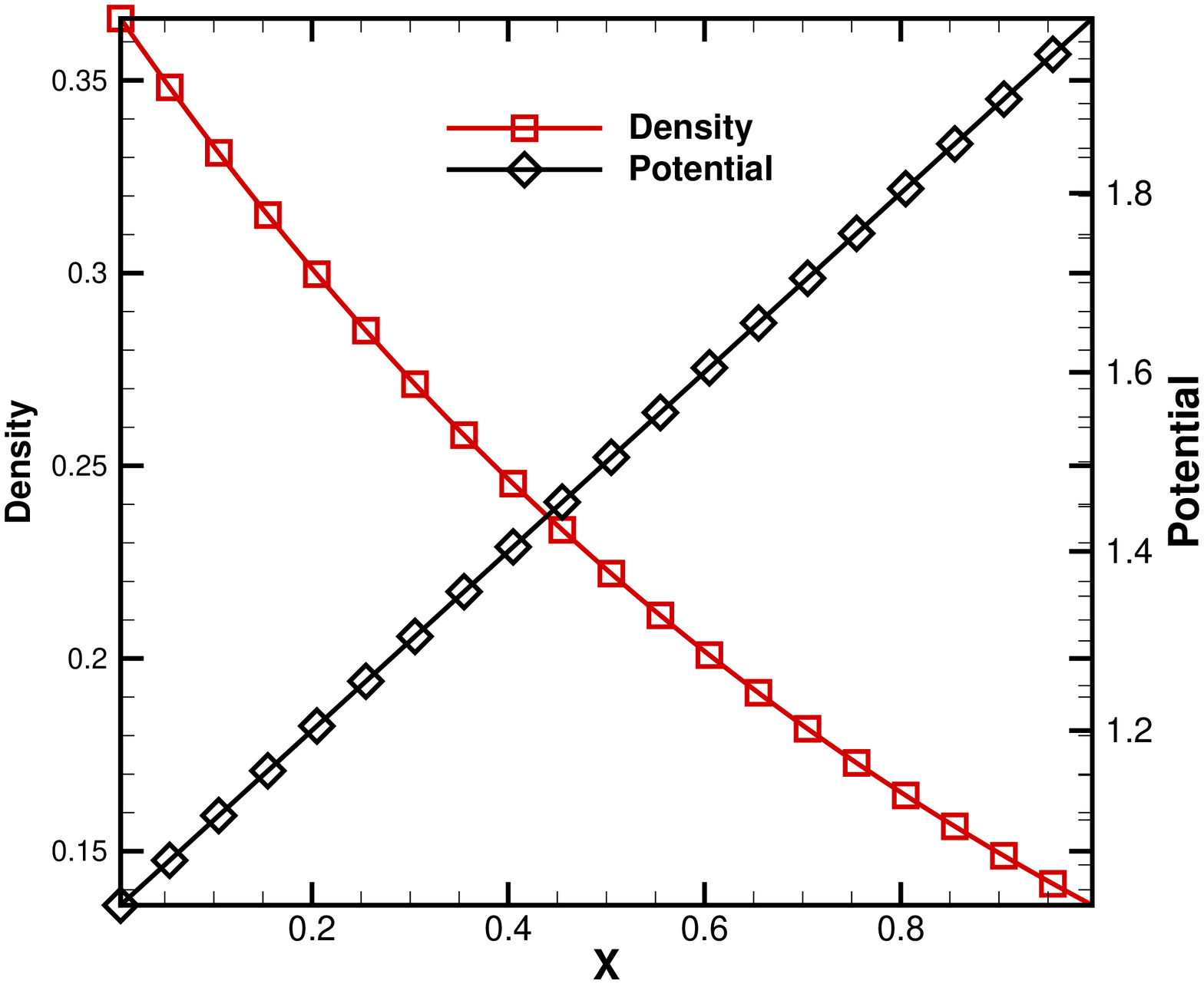}
    }
    \parbox[b]{0.31\textwidth}{
    (b)
    \includegraphics[totalheight=4.2cm, bb = 0 25 690 565, clip =
    true]{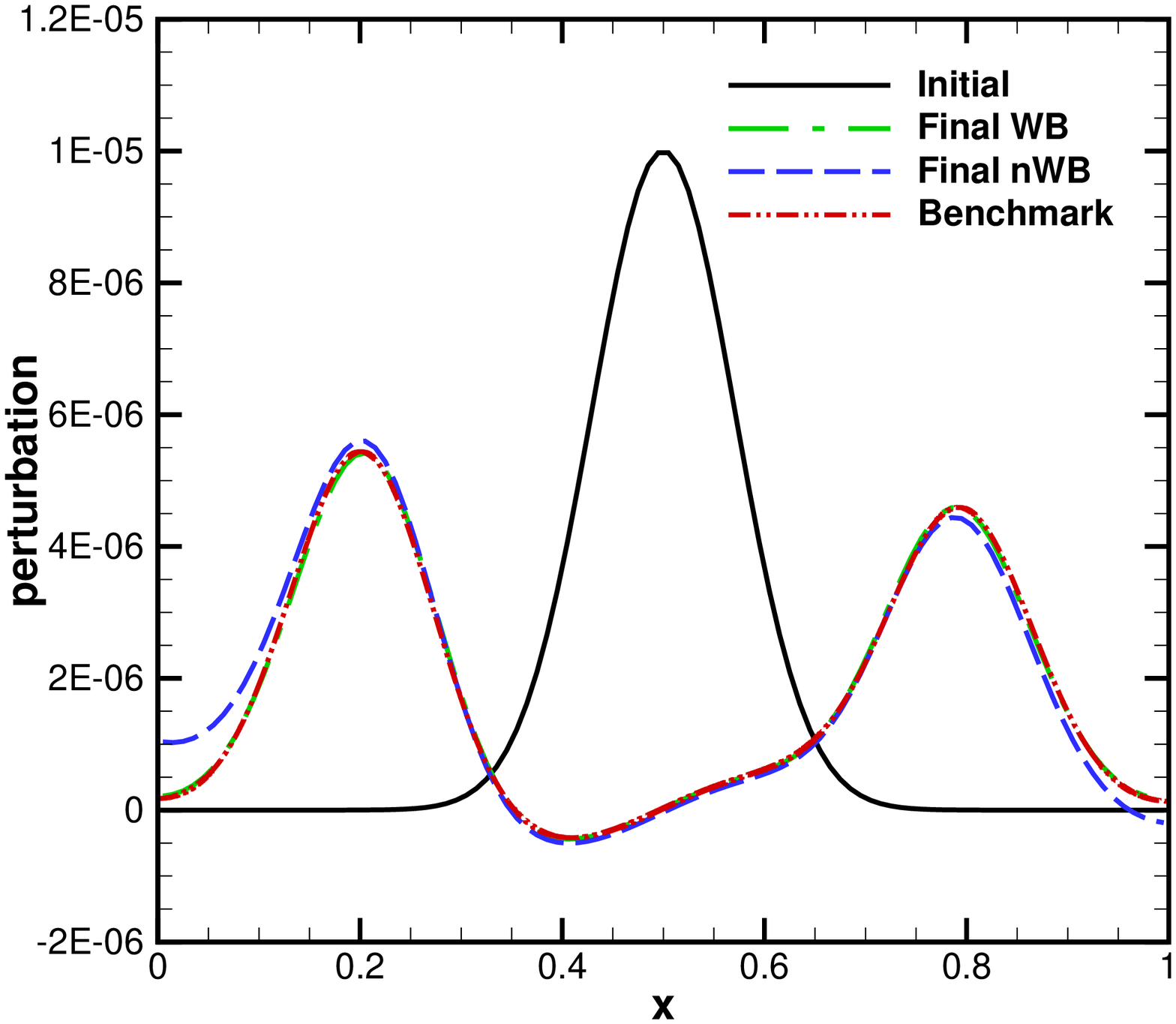}
    }
    \parbox[b]{0.31\textwidth}{
    (c)
    \includegraphics[totalheight=4.2cm, bb = 0 23 690 565, clip =
    true]{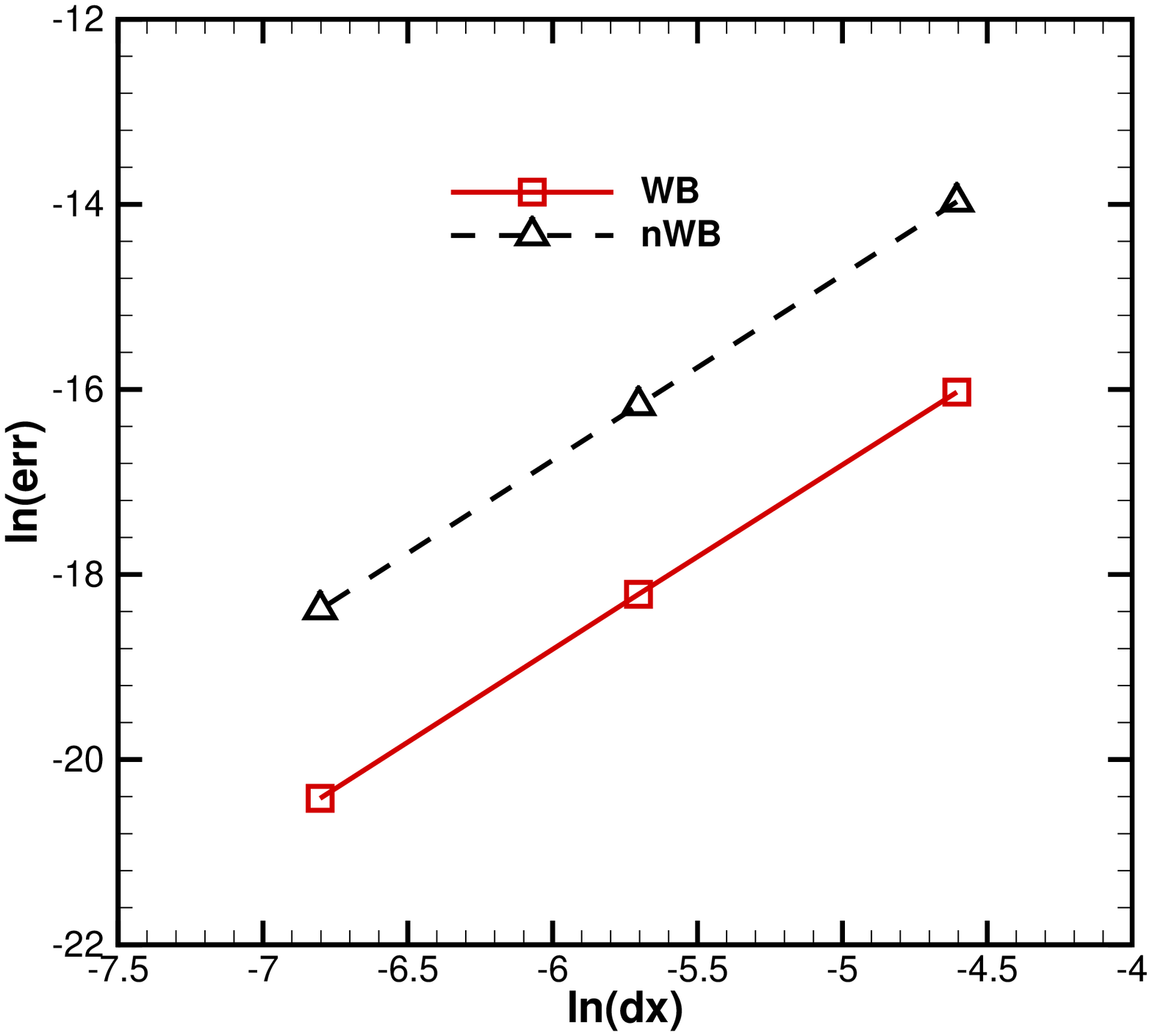}
    }
    \caption{The propagation of small perturbation ($\eta = 10^{-5}$) riding on an hydrostatic equilibrium state. (a) the initial condition; (b) the final solution; (c) the convergence rate.}
    \label{fig:perturbation-5}
\end{figure}

The Euler equations are considered firstly. The initial conditions are shown in Fig. \ref{fig:perturbation-5}(a). The pressure is disturbed by a moderate small perturbation, $\eta=10^{-5}$. The benchmark solution is derived by the present well-balanced scheme with 24300 uniform cells. As shown in the figure \ref{fig:perturbation-5}(b), the green dash dot line derived by WB scheme is very close to the benchmark solution, while the result derived by nWB scheme deviates from the benchmark solution. Then the convergence study is conducted by refining the mesh. And the error is defined as the $L_1$ norm of the deviation from the benchmark solution. Figure \ref{fig:perturbation-5}(c) shows that, the convergence rate is 1.9974 for the WB scheme, and is 2.0050 for the nWB scheme. Although both of them are of second order spatial accuracy, the WB scheme is much more accurate than the nWB scheme on coarse mesh.

\begin{figure}
\centering
    \parbox[b]{0.48\textwidth}{
    \includegraphics[totalheight=5.5cm, bb = 0 25 690 565, clip =
    true]{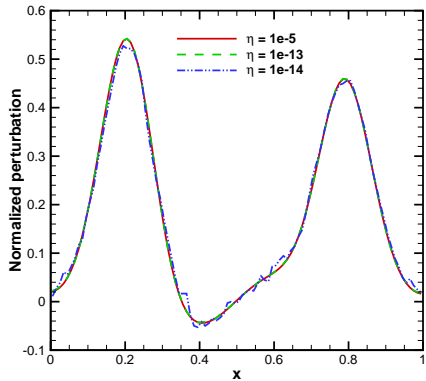}
    }
    \caption{The normalized solutions of the pressure perturbation on the isothermal equilibrium state. The solutions of $\eta = 10^{-5}, 10^{-13}$ collapsing to a single curve indicates the well-balanced property of the present scheme.}
    \label{fig:perturbation-all}
\end{figure}

Then we tested two smaller amplitudes of the perturbation, $\eta = 10^{-13}$ and $10^{-14}$. Since no other numerical results has been reported for such small perturbation, we propose a self-evaluation procedure to assess the present well-balanced scheme. As the perturbation is very small, the flow system will respond linearly, namely, if normalize the numerical results by their amplitude of the initial perturbation, the rescaled numerical results will collapse to a single curve. Figure \ref{fig:perturbation-all} shows that the normalized solutions for $\eta = 10^{-5}, 10^{-13}$ coincide with each other, but the normalized solution for $\eta = 10^{-14}$ deviates from others on the level of $10^{-16}$.
In fact, the rounding error is about $10^{-16}$ if double-precision is adopted in the computation.

\subsubsection{Viscous flow}
\begin{figure}
\centering
    \parbox[b]{0.48\textwidth}{
    \includegraphics[totalheight=5.5cm, bb = 0 25 690 565, clip =
    true]{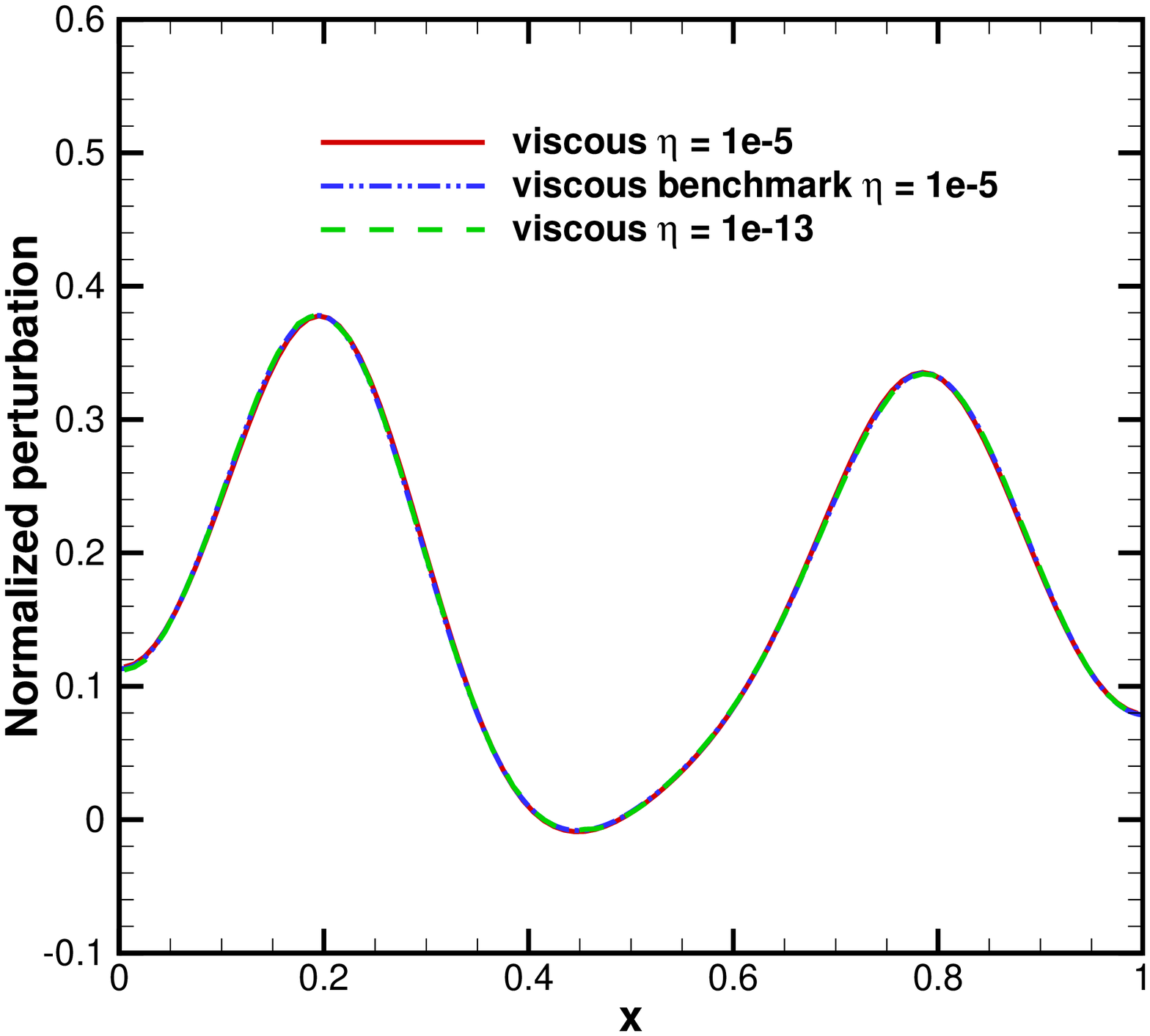}
    }
    \parbox[b]{0.48\textwidth}{
    \includegraphics[totalheight=5.5cm, bb = 0 23 690 565, clip =
    true]{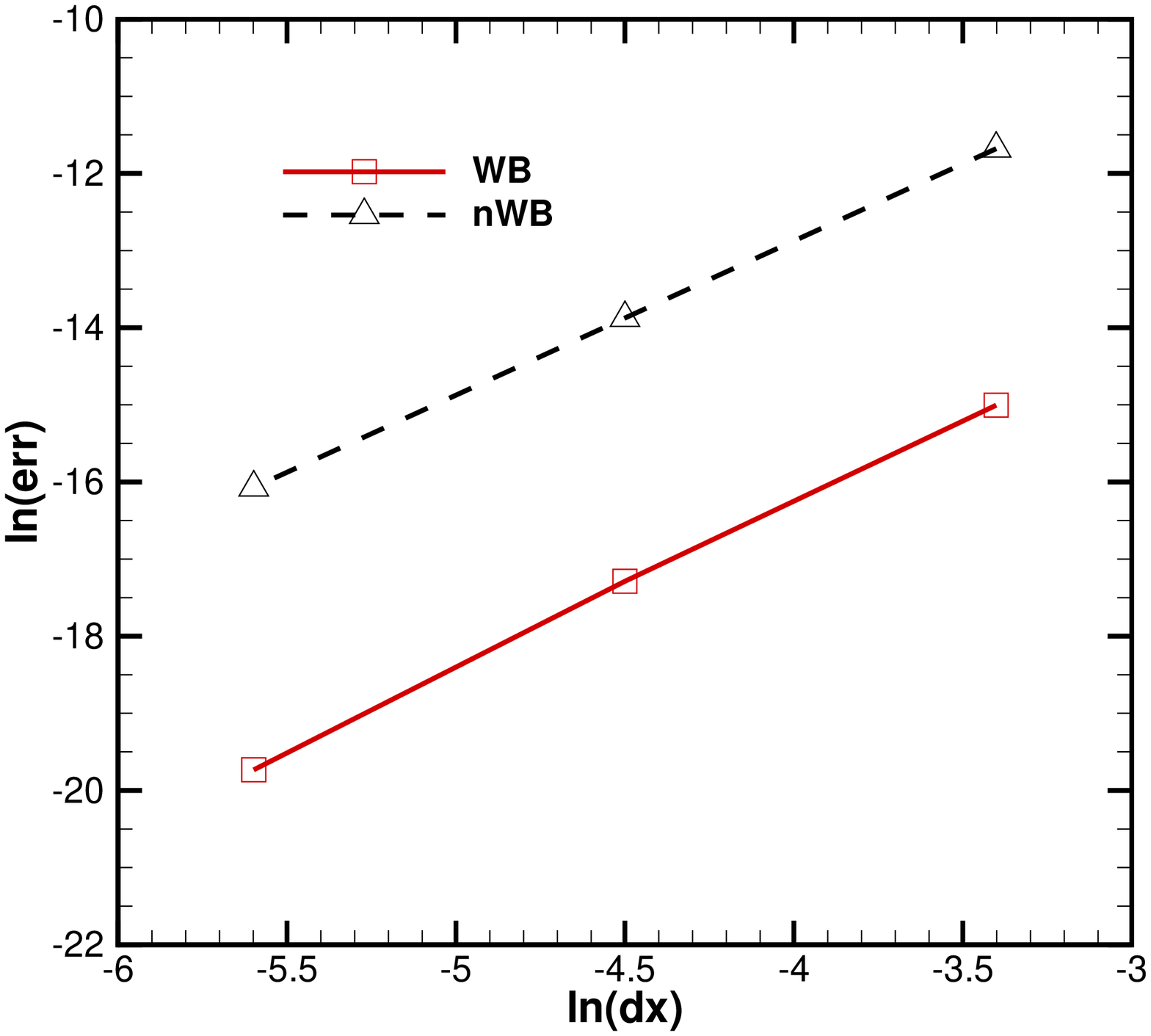}
    }
    \caption{The propagation of the pressure perturbation on the isothermal equilibrium state in viscous fluid ($\nu = 0.01$). Left: the perturbation; Right: the convergence rate. The benchmark solution is derived by the present well-balanced scheme with 7290 uniform cells.
    }
    \label{fig:perturbation-vis}
\end{figure}

As we claimed the well-balanced scheme for the Navier-Stokes equations, the small perturbation propagating in viscous flow ($\nu = 0.01$) is also simulated. Two amplitudes of the perturbation, $\eta = 10^{-5}, 10^{-13}$, are considered and the results are also compared with the benchmark solution derived by the present WB scheme with 7290 uniform cells. As shown in Fig. \ref{fig:perturbation-vis}, the final amplitudes are smaller than that in the inviscid fluid. More importantly, the normalized solutions are also identical to each other, which means the present scheme for viscous term discretization is well-balanced. Otherwise, the truncation error will pollute the numerical results for smaller perturbation more severely, and separate two profiles in Fig.\ref{fig:perturbation-vis}. Similar to the inviscid case, the convergence rate is 2.1519 for the WB scheme, and is 1.9995 for the nWB scheme.
This challenging test demonstrated that our scheme can predict very accurate numerical results for small perturbations up to the machine zero for the Navier-Stokes equations.

\subsection{Evolution towards the isothermal hydrostatic equilibrium state}

Since the Euler equations possess many equilibrium states, a sort of well-balanced scheme has been proposed for a variety of equilibrium states \cite{chandrashekar2015second}. On the other hand, without heat conduction, the fluid system governed by the Euler equations allows temperature stratification, and cannot determine which equilibrium state the system will eventually stay at. As a result, the previous Euler equations' well-balanced schemes generally did not test the convergence towards isothermal hydrostatic equilibrium state. As a contrast, the Navier-Stokes equations only allow the fluid system eventually converge to the isothermal hydrostatic equilibrium state for an isolated system. Therefore, the evolution towards the isothermal hydrostatic equilibrium state distinguishes the well-balanced scheme for the Navier-Stokes equations from the well-balanced scheme for the Euler equations.
\begin{figure}
    \raisebox{2.1cm}{$\phi_1 = x+1$} \hfill
    \parbox[b]{0.4\textwidth}{
    t=40
    \includegraphics[totalheight=4.5cm,bb = 0 26 690 525, clip =
    true]{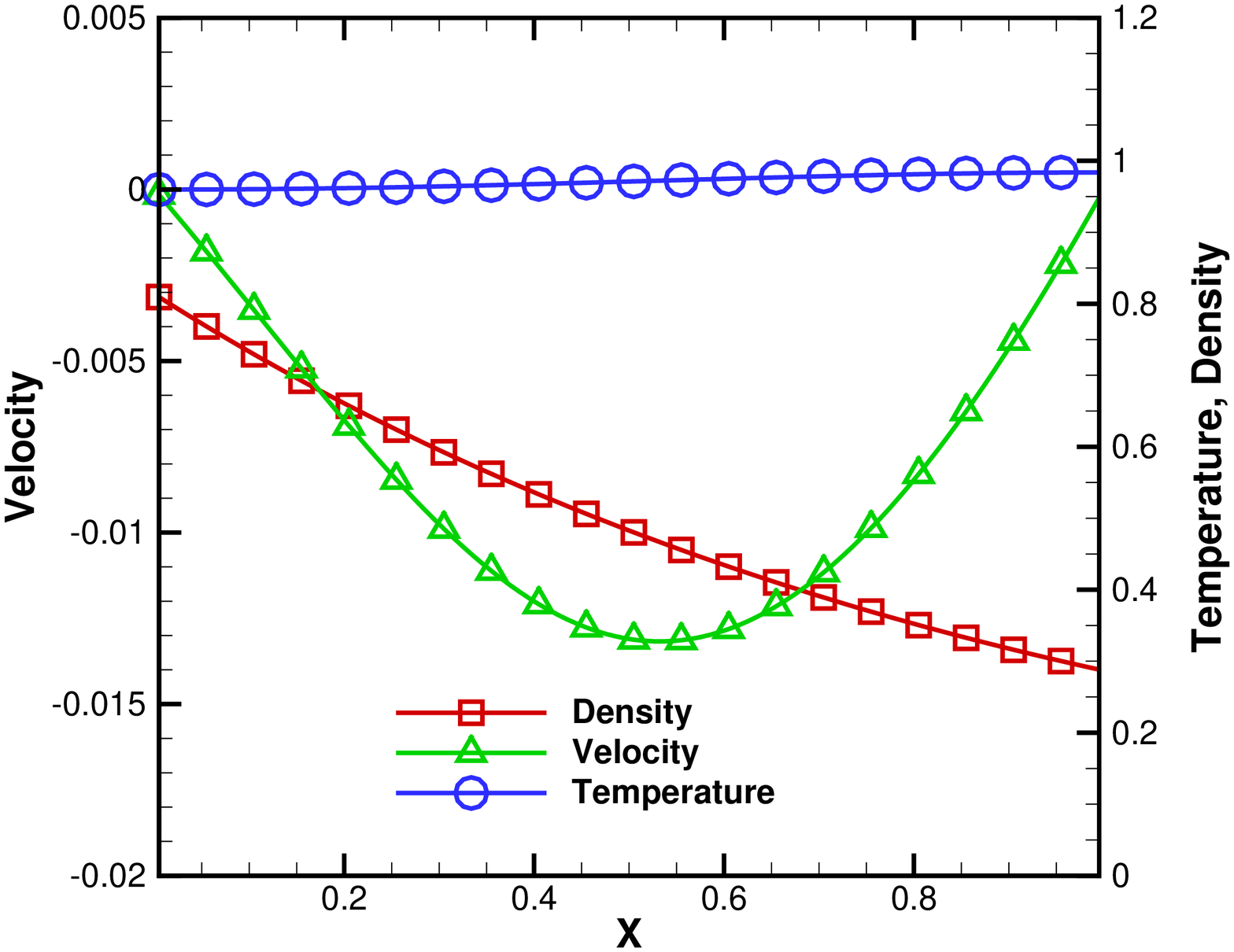}
    }
    \parbox[b]{0.4\textwidth}{
    t=1000
    \includegraphics[totalheight=4.5cm,bb = 0 26 690 525, clip =
    true]{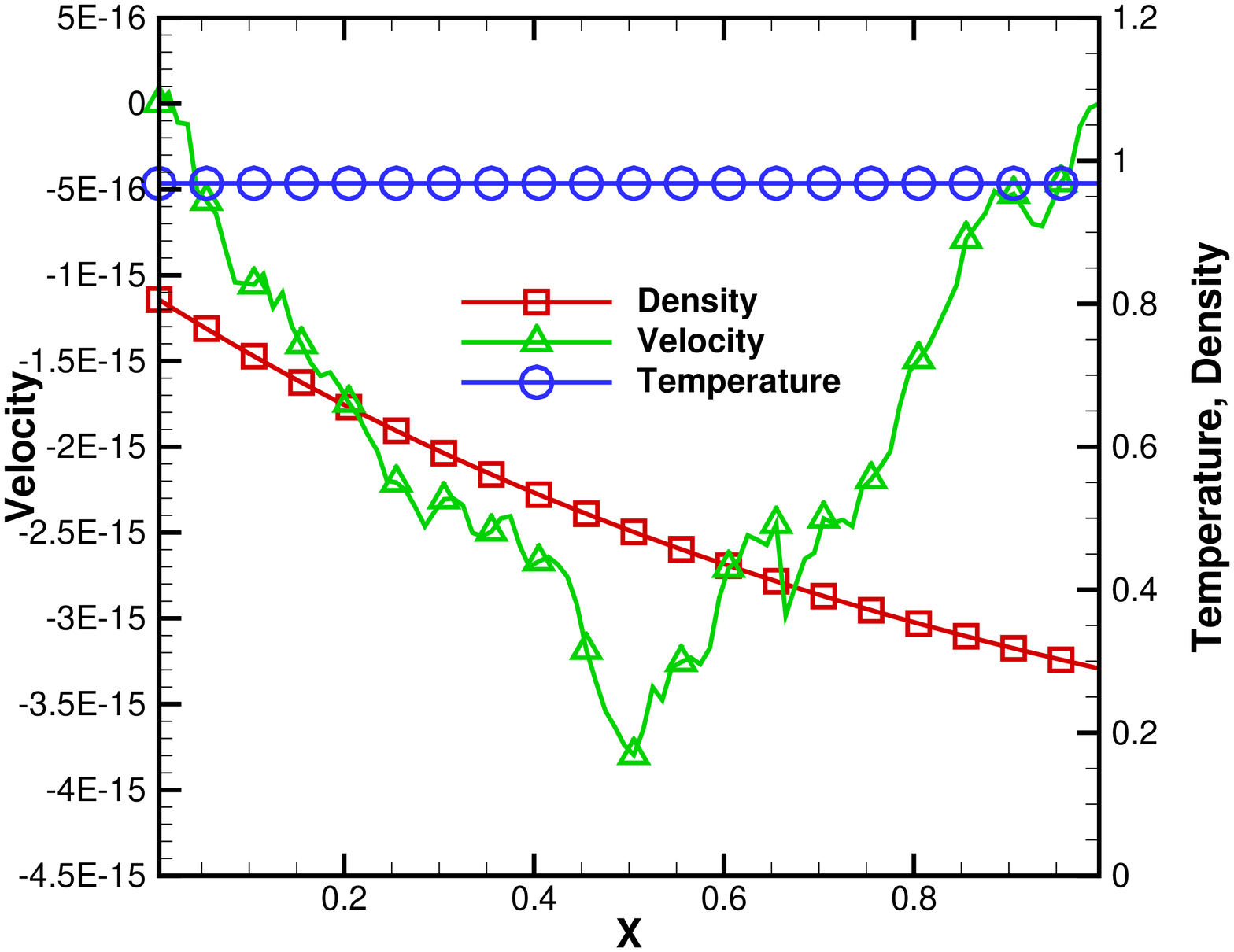}
    }
    \raisebox{2.1cm}{$\phi_2 = x^2+1$} \hfill
    \parbox[b]{0.4\textwidth}{
    \includegraphics[totalheight=4.5cm,bb = 0 26 690 525, clip =
    true]{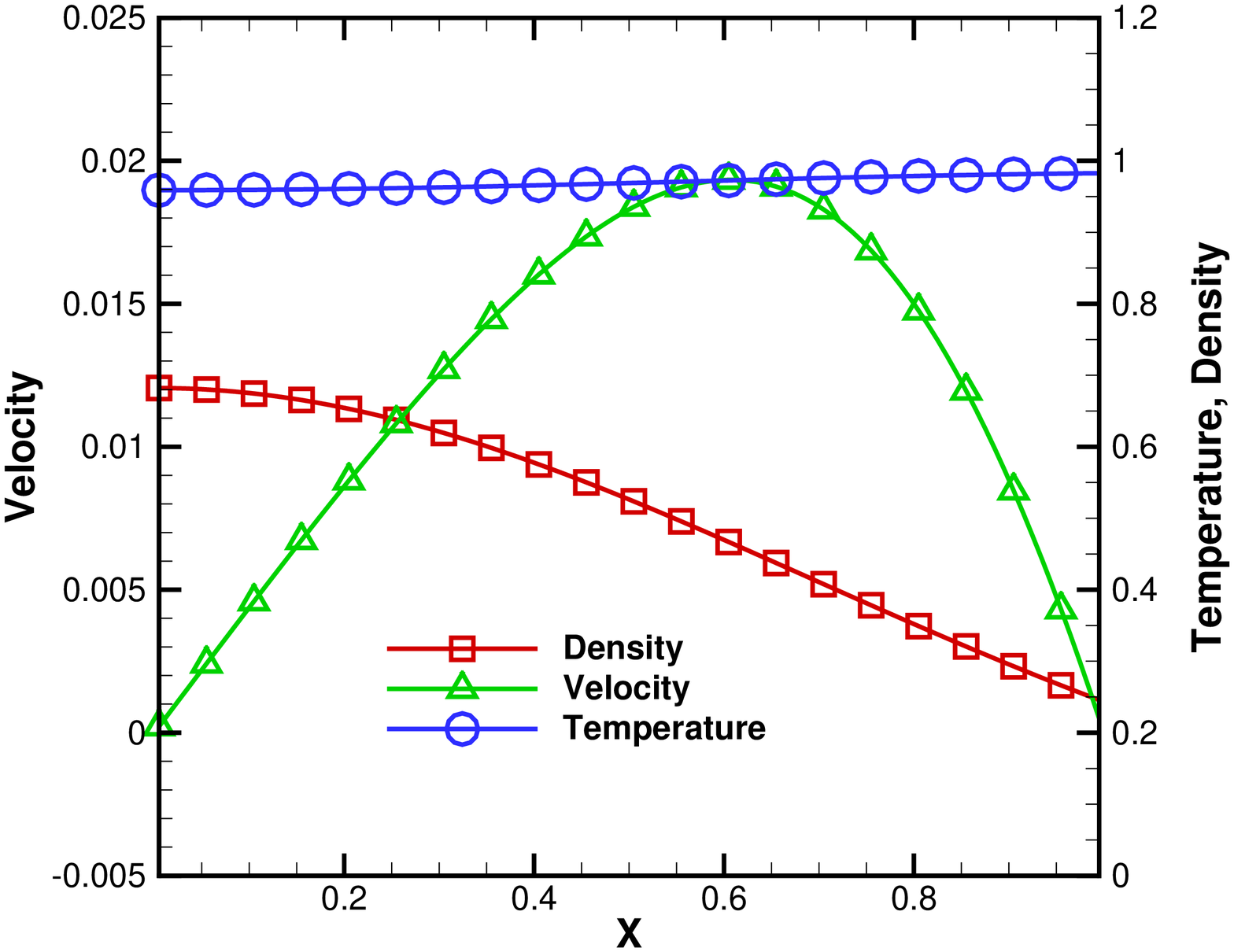}
    }
    \parbox[b]{0.4\textwidth}{
    \includegraphics[totalheight=4.5cm,bb = 0 26 690 525, clip =
    true]{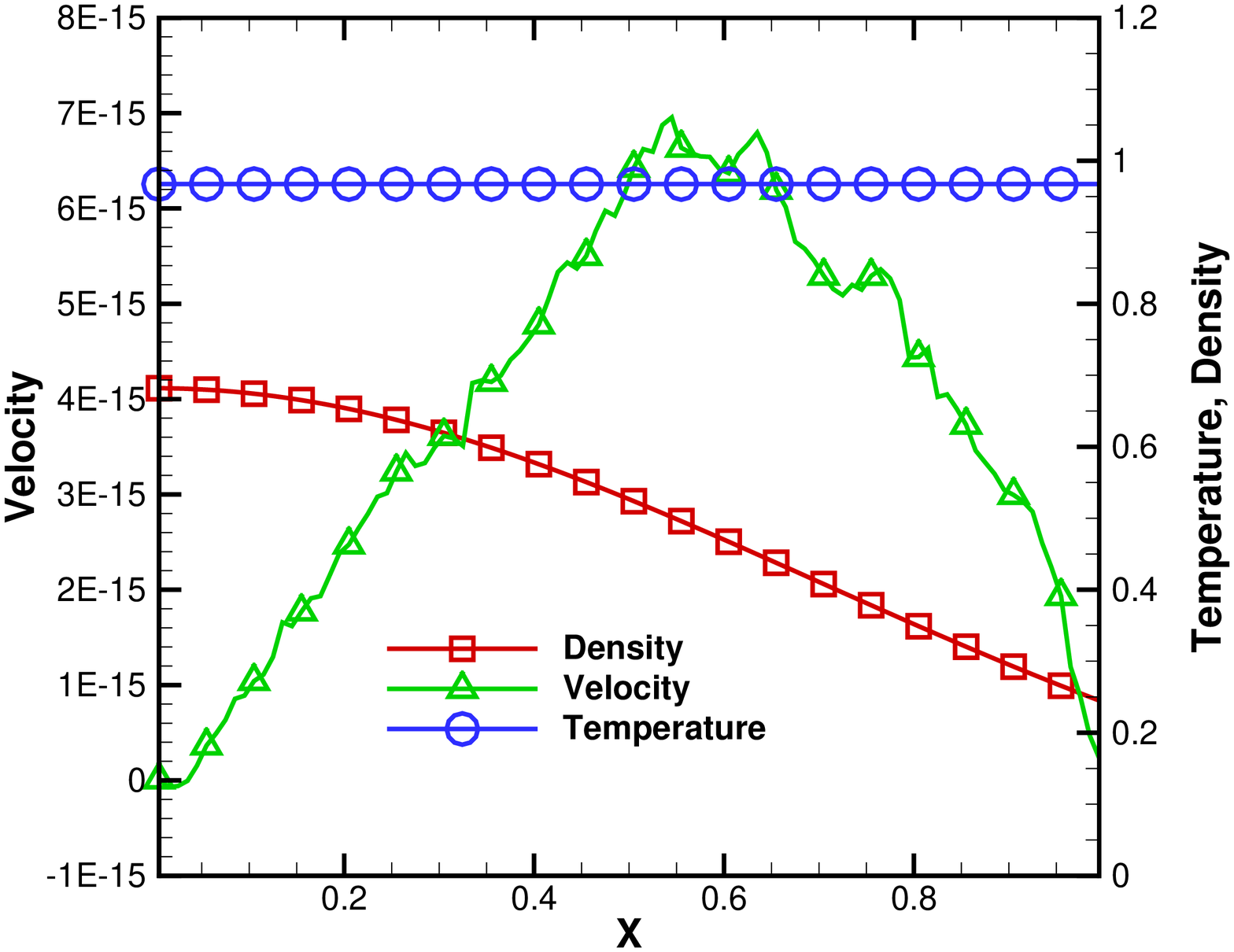}
    }
    \raisebox{2.1cm}{$\phi_3 = \sin(2\pi x)+2$} \hfill
    \parbox[b]{0.4\textwidth}{
    \includegraphics[totalheight=4.5cm,bb = 0 26 690 525, clip =
    true]{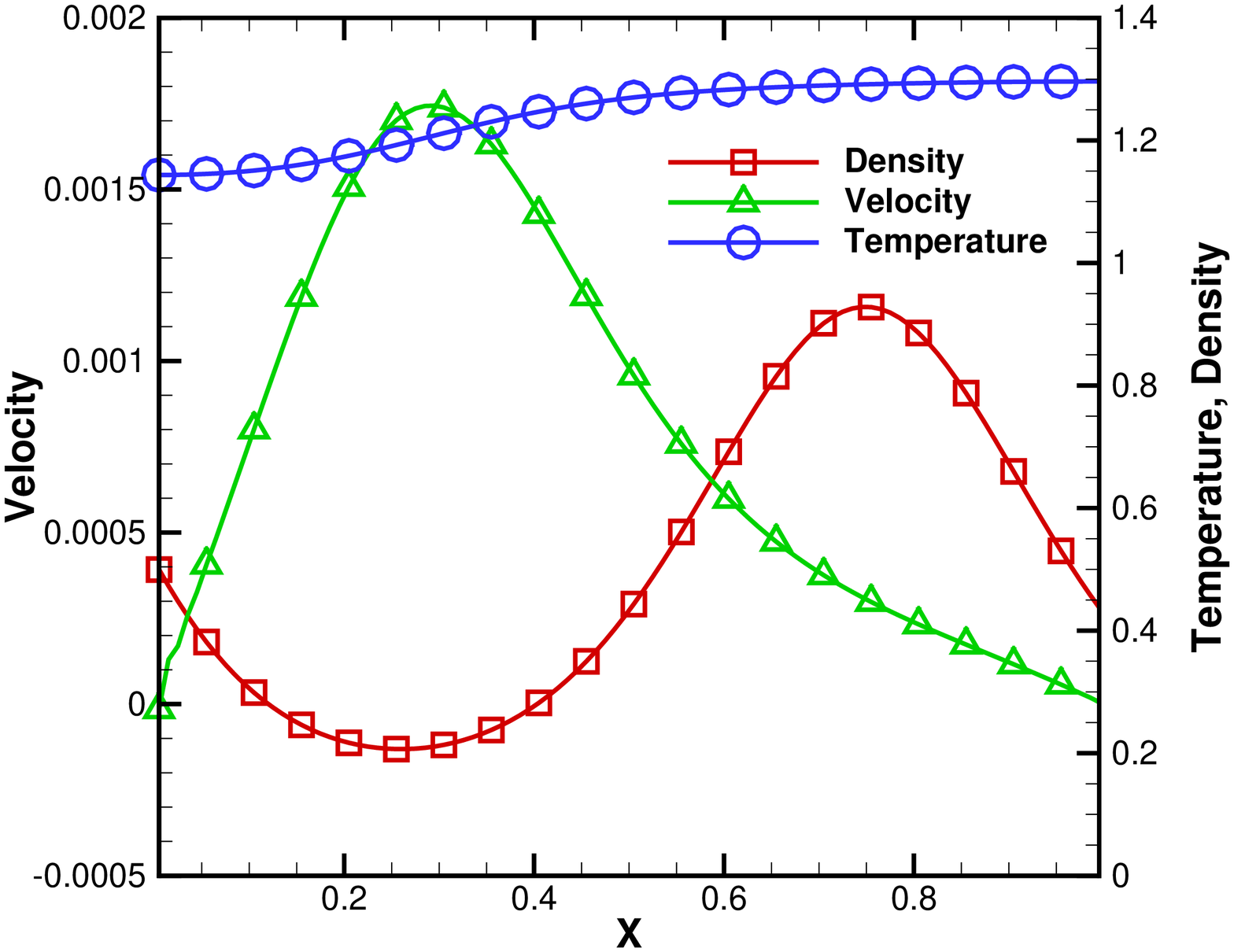}
    }
    \parbox[b]{0.4\textwidth}{
    \includegraphics[totalheight=4.5cm,bb = 0 26 690 525, clip =
    true]{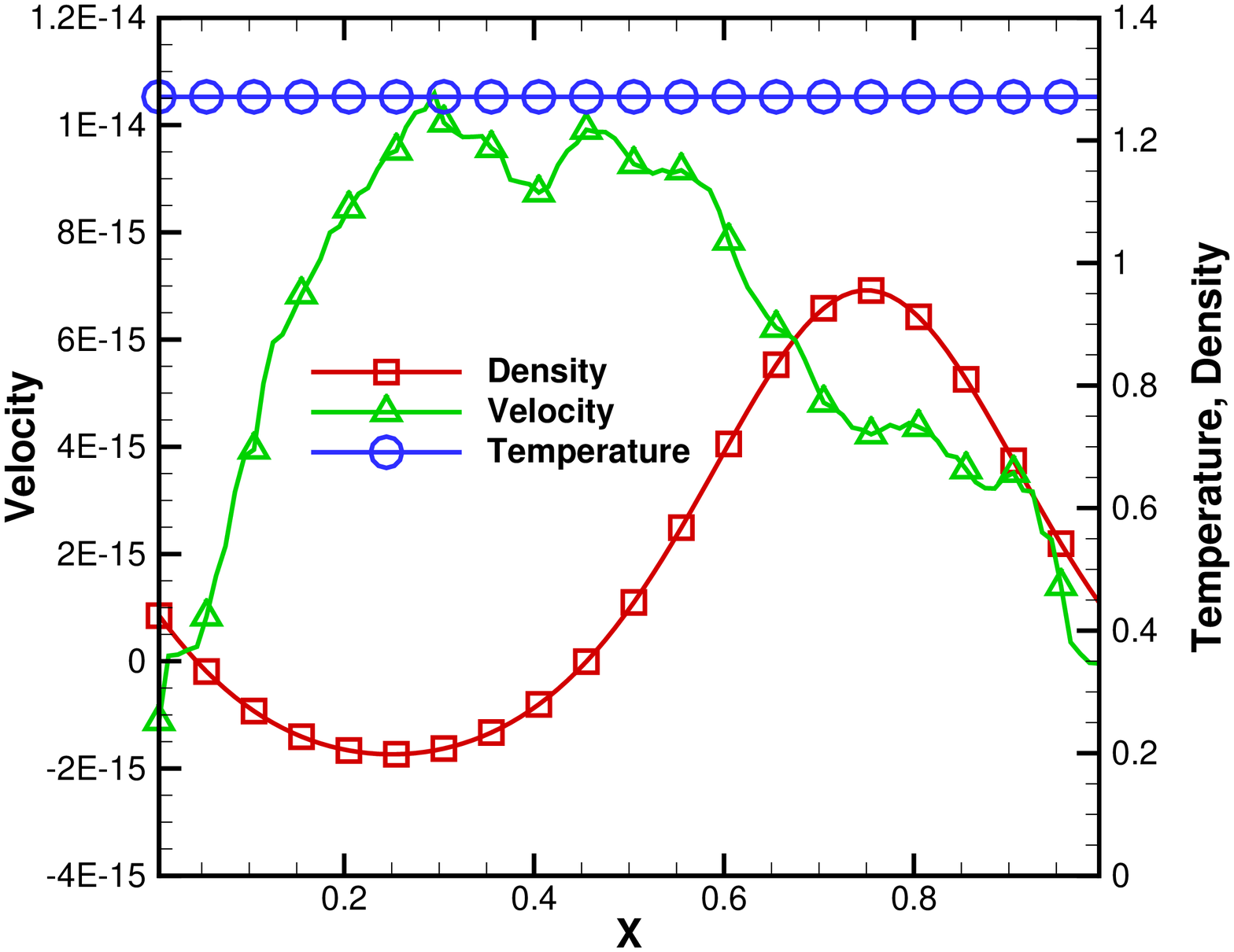}
    }
    \caption{Evolution towards the isothermal hydrostatic equilibrium state under different potential functions.}
    \label{fig:evolve}
\end{figure}

To test the convergence process, the three potential functions (Eq.(\ref{eq:3-potentials})) are employed again, and the gas properties are $\nu=0.005$, $\gamma = 1.4$ and $\text{Pr} = 1$. The initial condition is given as follows,
\begin{eqnarray}
\rho = 1-x, \quad U = 0, \quad T = 1, \quad x\in(0,1).
\end{eqnarray}
Notice that this initial condition deviates far from the equilibrium state regarding anyone of the three potential functions.
The computational domain is uniformly divided into 100 cells. The simulations stop at $t = 1000$.
Fig. \ref{fig:evolve} shows the velocity and temperature solution at $t = 40,1000$ under the three potential functions. It can be seen that the system converges to the stationary isothermal hydrostatic equilibrium state for all three potential functions. Even for the most complex sine function, the final velocity is less than $10^{-14}$.
\begin{figure}
\centering
    \parbox[b]{0.32\textwidth}{
    $t=200$
    \includegraphics[totalheight=4.2cm,bb = 0 26 690 525, clip =
    true]{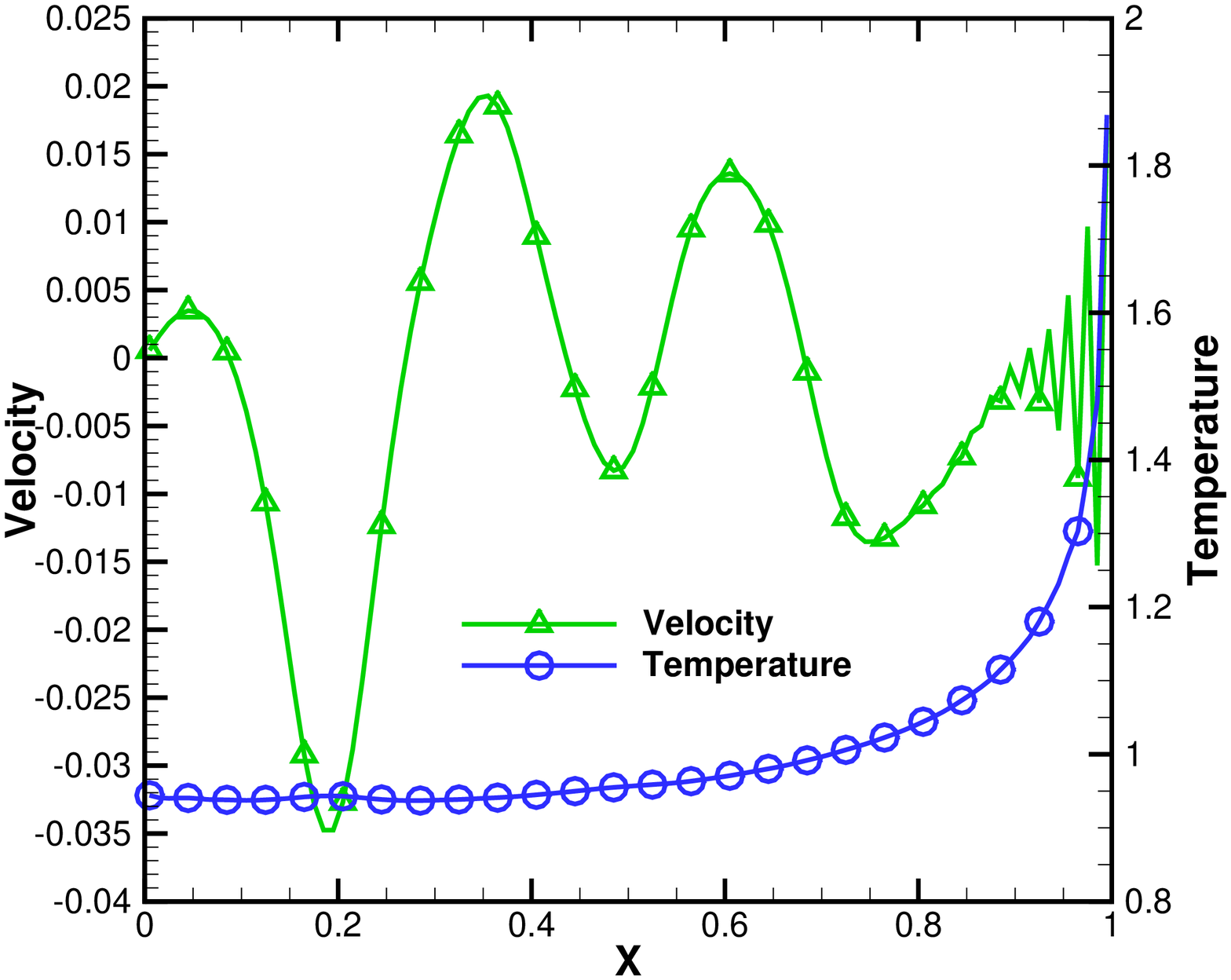}
    }
    \parbox[b]{0.32\textwidth}{
    $t=600$
    \includegraphics[totalheight=4.2cm,bb = 0 26 690 525, clip =
    true]{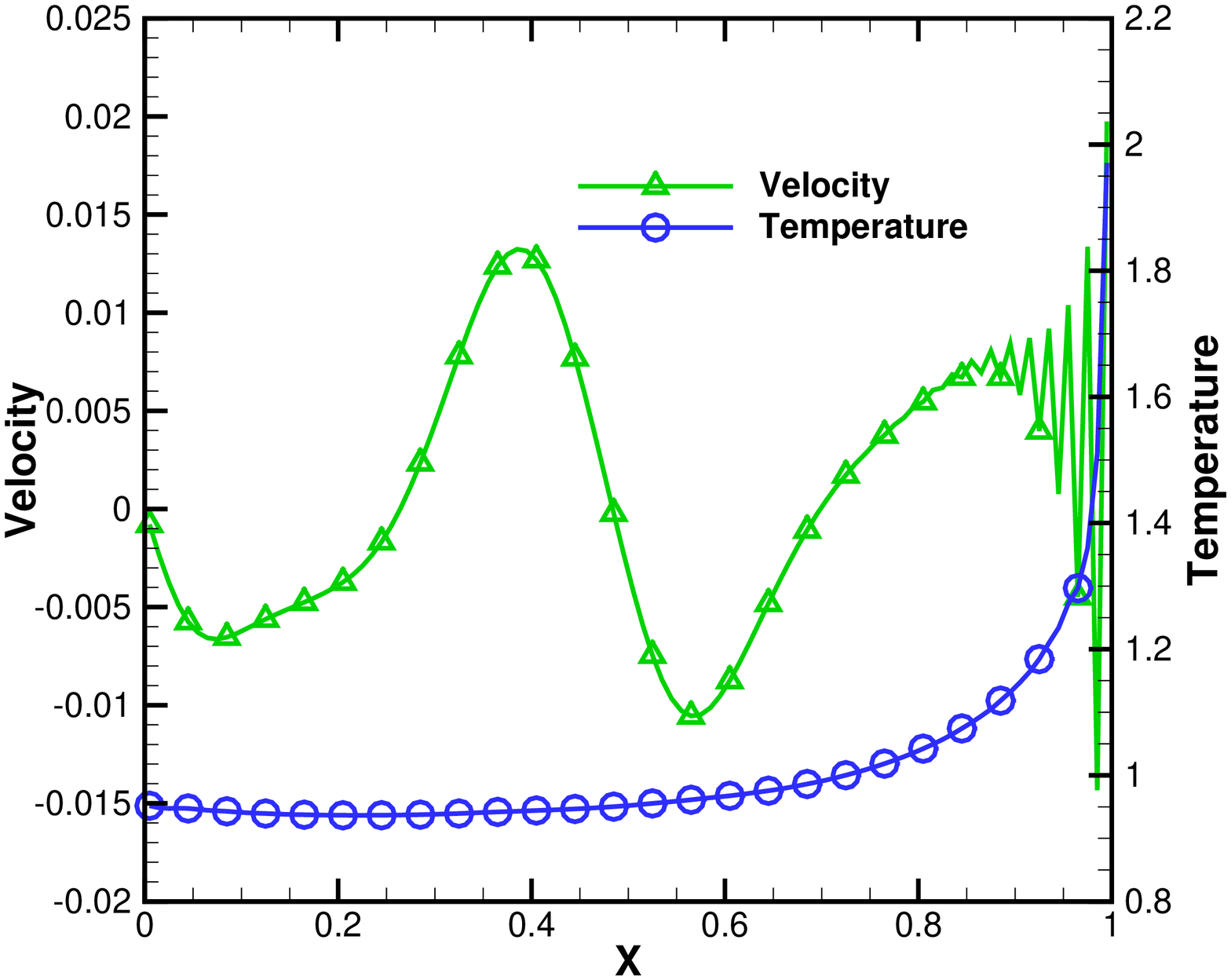}
    }
    \parbox[b]{0.32\textwidth}{
    $t=1000$
    \includegraphics[totalheight=4.2cm,bb = 0 26 690 525, clip =
    true]{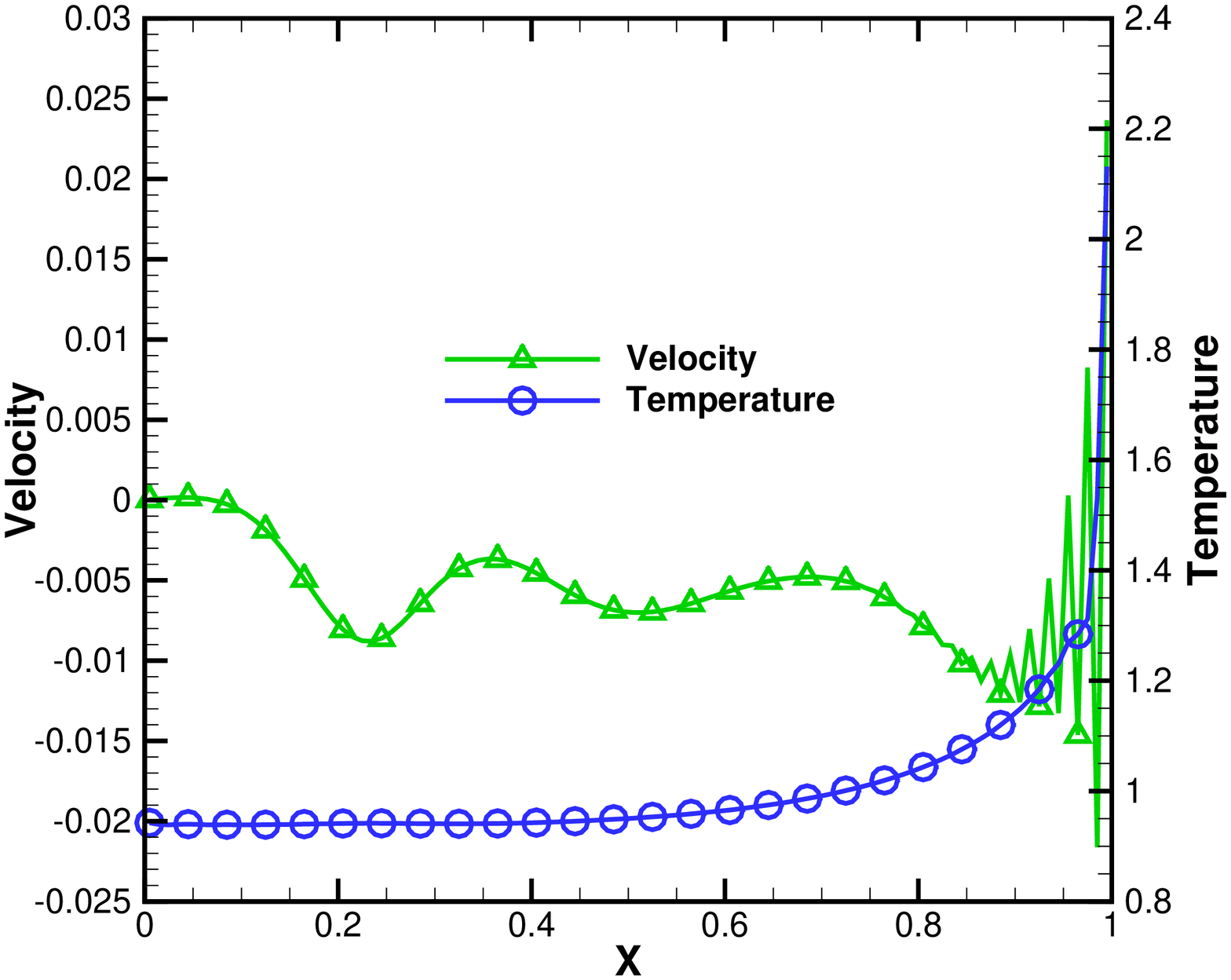}
    }
    \caption{Evolution towards the isothermal hydrostatic equilibrium state under linear potential function ($\phi=x+1$) with zero viscosity and zero thermal conductivity.}
    \label{fig:evolveEuler}
\end{figure}

By contrast, Fig. \ref{fig:evolveEuler} shows gas evolution at three different times with $\phi_1$, zero viscosity and zero thermal conductivity. It can be seen that the vital movement of gas lasts for long time. Since the light gas with high temperature will move towards the end of low potential, the high temperature gas will accumulate at the right end if the thermal conduction is absent. As a result, the temperature profile becomes very steep, and numerical solution oscillates due to the central difference in the scheme. Fig. \ref{fig:evolveEuler} indicates that the Euler equations are inappropriate for long time simulation because of the lack of the physical dissipation mechanism.

\subsection{2D Rayleigh-Taylor instability}
\label{sec:RTinstability}
In this subsection, two dimensional two-layer flow under gravity is simulated to validate the proposed scheme.
A radially symmetric potential is given in polar coordinate as follows,
\begin{eqnarray}
\phi(r,\theta) = ar + b,
\end{eqnarray}
where $r$ is the radius, and $\theta$ is the azimuth in polar coordinate.
Two isothermal equilibrium states are assigned to the inner layer ($r < r_i$) and outer layer ($r\ge r_i$) respectively,
\begin{eqnarray}
\left\{\begin{array}{ccl}
\rho(r) &=& \rho_0 e^{-\frac{a(r-r_0)}{RT_0}} \\
p(r) &=& \rho(r) RT_0
\end{array}\right., \quad \quad r < r_i;\label{eq:RTinner}
\end{eqnarray}
\begin{eqnarray}
\left\{\begin{array}{ccl}
\rho(r) &=& (\rho_0+\Delta\rho) e^{-\frac{a(r-r_0)}{RT_1}} \\
p(r) &=& \rho(r) RT_1 \\
T_1 &=& T_0\frac{\rho_0}{\rho_0+\Delta\rho}
\end{array}\right., \quad \quad r \geq r_i.\label{eq:RTouter}
\end{eqnarray}
where $\rho_0 = 0.1$, $\Delta \rho = 0.1$, $a = 1.0$, $b = 1.0$, $T_0 = 0.3$ and $R = 1.0$.
An interface located at $r = r_i$ separates the two layers, and is twisted so as to make the cold fluid (denser at the interface) penetrate into the hot fluid (lighter at the interface),
\begin{eqnarray}
r_i = r_0+(1+\eta\cos(\kappa\theta)). \label{eq:RTinterface}
\end{eqnarray}
where $r_0 = 0.6$, $\eta = 0.02$, $\kappa = 20$.
Note that, there are varied numerical configurations in previous literature \cite{leveque1999wave,tian2007gravity,luo2011wb,chandrashekar2015second}.
In the following numerical simulations, the simpler configuration (Eqs.(\ref{eq:RTinner},\ref{eq:RTouter},\ref{eq:RTinterface})) is adopted and a complete set of parameter including kinematic viscosity is presented. The simulations are conducted on a uniform grid covering $[-1,1]\times[-1,1]$ with $200\times 200$ cells.

\begin{figure}
\centering
    \parbox[b]{0.48\textwidth}{
    (a) $\nu = 0.0001$
    \includegraphics[totalheight=5.5cm]{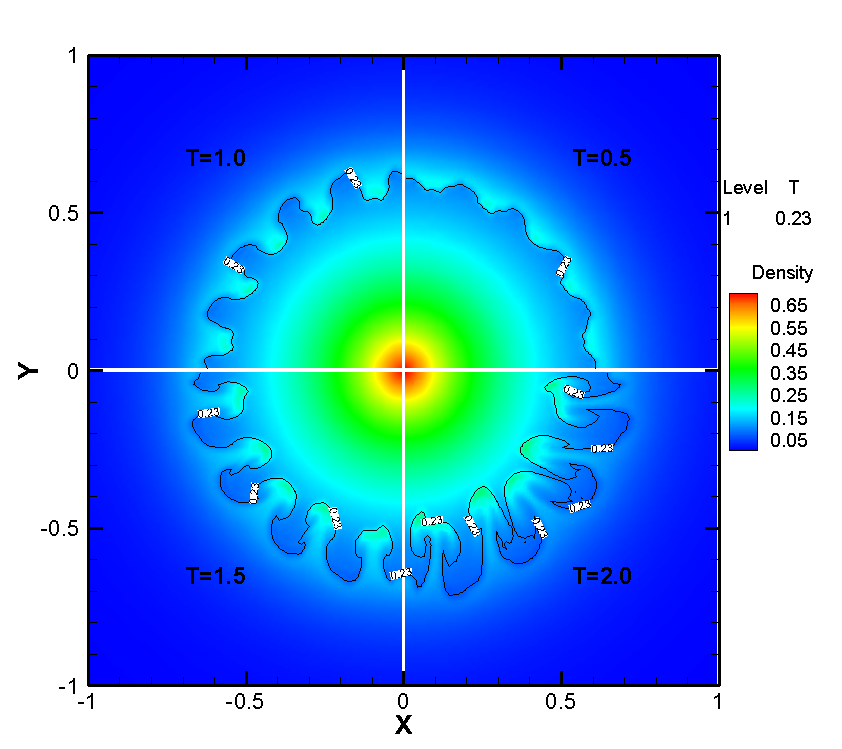}
    }
    \parbox[b]{0.48\textwidth}{
    (b) $\nu = 0.0002$
    \includegraphics[totalheight=5.5cm]{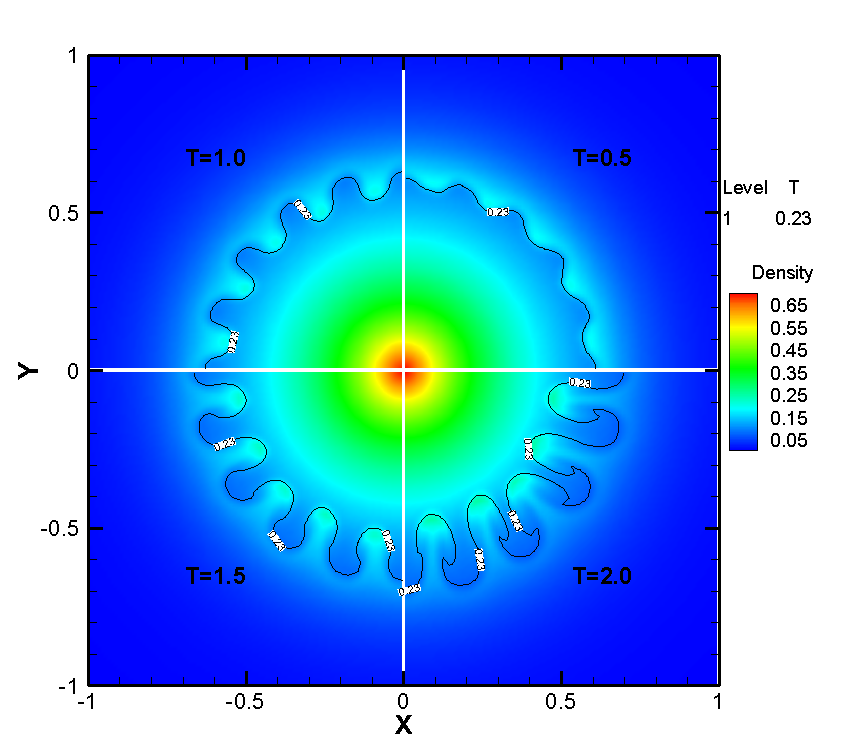}
    }
    \parbox[b]{0.48\textwidth}{
    (c) $\nu = 0.0004$
    \includegraphics[totalheight=5.5cm]{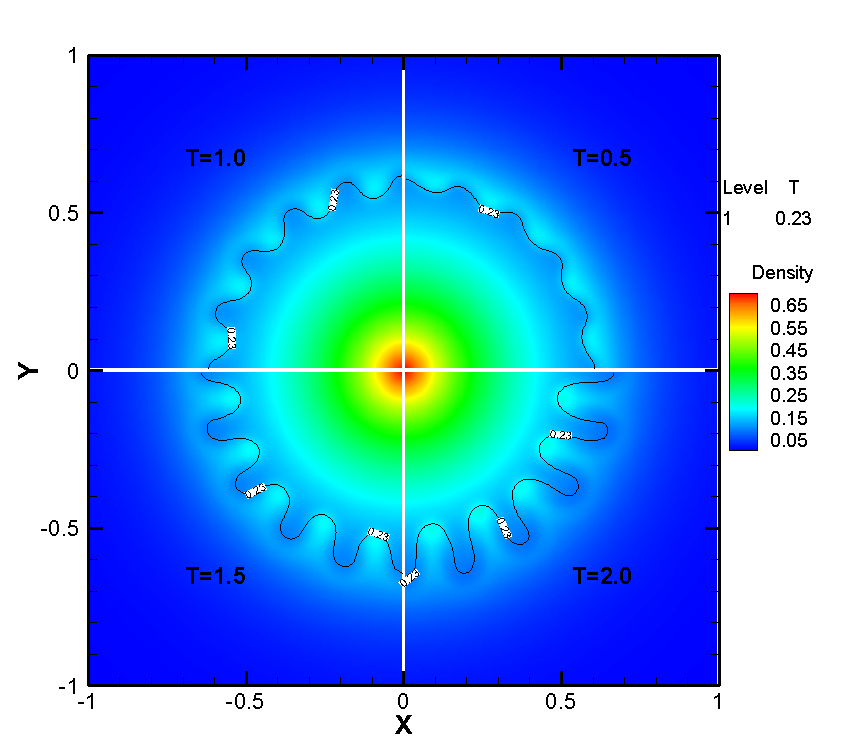}
    }
    \parbox[b]{0.48\textwidth}{
    (d) $\nu = 0.0008$
    \includegraphics[totalheight=5.5cm]{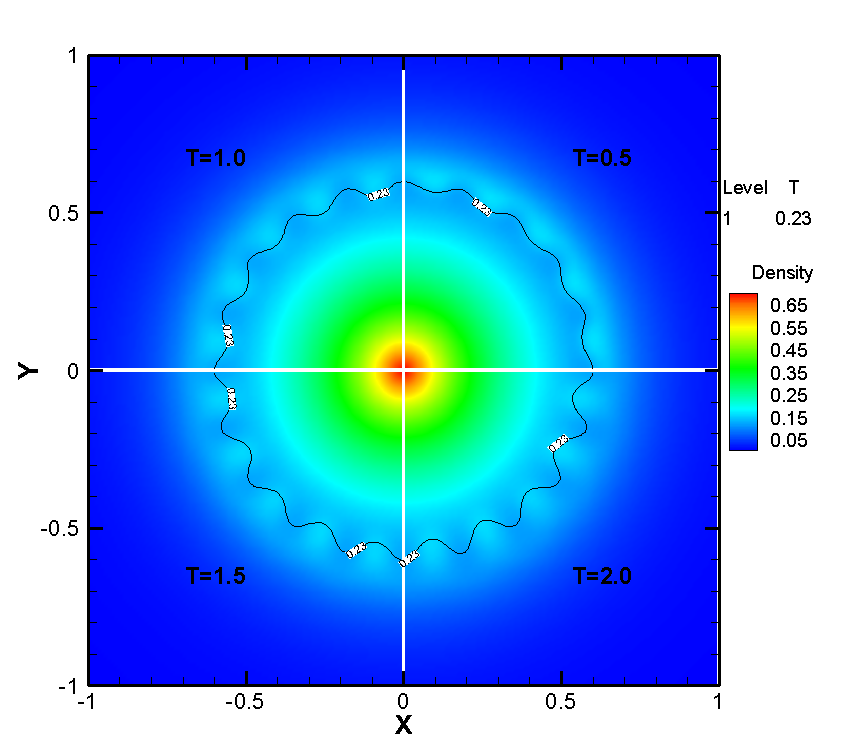}
    }
    \caption{The flow pattern of two dimensional Rayleigh-Taylor instability with different kinematic viscosity ($\nu$) at $t = 0.5, 1.0, 1.5, 2.0$. Small $\nu$ (a) distorts the solution as the Cartesian mesh breaks the symmetry; moderate $\nu$ (b) stabilizes the spikes and maintains the symmetry; large $\nu$ (c,d) smoothes the mushroom-shaped structure; }
    \label{fig:visRT}
\end{figure}

In Fig. \ref{fig:visRT}, the density contour is presented and a temperature contour line ($T=0.23$) which separates the hot and cold fluids is also provided to illustrate the spike structure more clearly.
Theoretically, if polar grid system is employed, the numerical solution will be symmetric. However, the projection onto the Cartesian mesh can be regarded as a force disturbing the symmetry. This is the origin of asymmetrical pattern in this test problem.
On the other hand, the viscous term can be regarded as a resistance to the disturbing force.
As shown in Fig. \ref{fig:visRT}, the flow pattern of the RT instability is highly dependent on the viscosity. When $\nu$ is small (Fig.(\ref{fig:visRT}(a)), the disturbance from the projection onto the Cartesian mesh breaks the symmetry of the solution. As $\nu$ increases (Fig.(\ref{fig:visRT}(b))), the viscous effect suppresses the disturbance and stabilizes the mushroom-shaped spikes and maintains the symmetry much better. Further increasing $\nu$ (Fig.(\ref{fig:visRT}(c,d))), the mushroom-shaped structures become smooth and further fade away with large viscosity. In fact, all these three types of flow pattern were reported in the literature\cite{leveque1999wave,tian2007gravity,luo2011wb,chandrashekar2015second}, because the numerical dissipation was implicitly implemented in their solvers of the Euler equations.

\begin{figure}
\centering
    \parbox[b]{0.6\textwidth}{
    \includegraphics[totalheight=8cm]{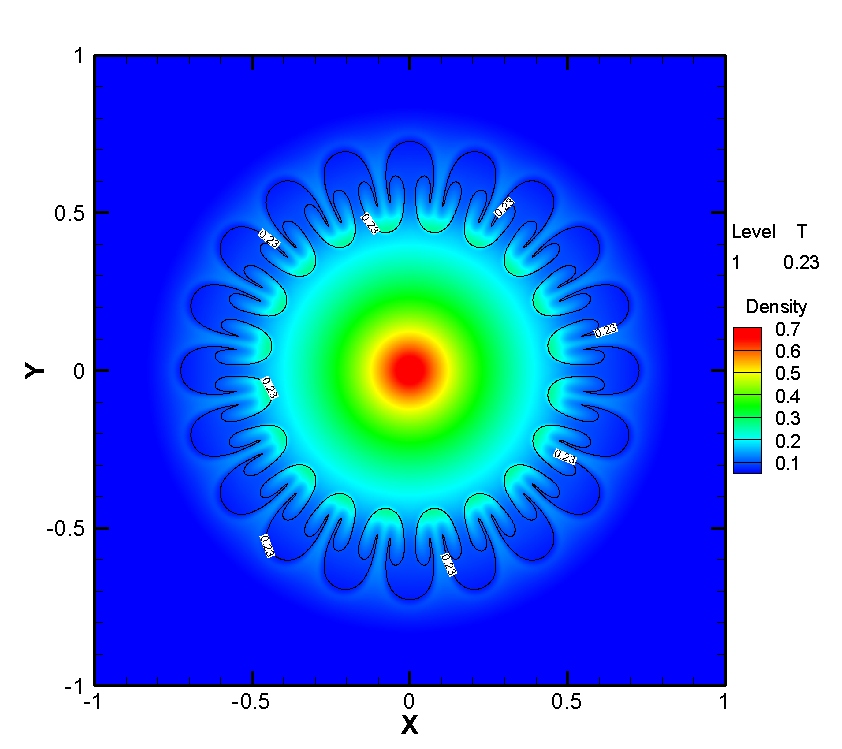}
    }
    \caption{The flow pattern of two dimensional Rayleigh-Taylor instability using $800\times 800$ cells at $t = 2.0, \nu = 0.0001$.} \label{fig:visRT800}
\end{figure}

We further refined the meshes, and present a high resolution result (Fig.(\ref{fig:visRT800})) on a $800 \times 800$ uniform mesh. It can be found that, since the disturbing force becomes smaller on the refined mesh, the flow pattern keeps its symmetry more easily even with small viscosity ($\nu = 0.0001$). According to this result, the numerical solutions on $200\times 200$ mesh are far from convergency.

\section{Conclusion}
In this study, we introduced an auxiliary variable which becomes constant at isothermal hydrostatic equilibrium state and reformulated the source term in the Navier-Stokes equations into a convenient form. A part of the original source term is merged into the flux term, and the remaining source term becomes zero at isothermal hydrostatic equilibrium state.

Based on the reformulated source term, we proposed a second-order well-balanced gas kinetic scheme for the Navier-Stokes equations. Through the global reconstruction of the auxiliary variables, the numerical fluxes and numerical source term vanish simultaneously when approaching the hydrostatic equilibrium state, which guarantees the well-balanced property. The new scheme has no assumption of the potential function, and hence, is simple and computationally efficient compared to symplecticity-preserving GKS.

Several test cases were presented to demonstrate the accuracy and the stability of the new scheme.
The one-dimensional hydrostatic equilibrium can be exactly held up to machine accuracy by the proposed scheme. The results for wave propagation riding on a hydrostatic equilibria has shown a significant gain in accuracy with current scheme. The small perturbation only several orders greater than the machine accuracy still survived as the simulation proceeded. Moreover, the linear response with viscous effect was also predicted accurately.

Another featured property of well-balanced scheme for the NS equations is the capability of simulating the evolution towards the hydrostatic equilibrium state. It requires that the physical dissipation and heat transfer, which are missing in the Euler equations, must be properly represented in the scheme for long period simulation.
And the proposed scheme correctly predicted an isothermal hydrostatic equilibrium state after a long running time.

In summary, a well-balanced gas kinetic scheme for the Navier-Stokes equations is proposed and validated through several challenging numerical problems. The evolution towards the isothermal hydrostatic equilibrium from highly non-equilibrium state is simulated and the final equilibria is achieved. The current scheme is capable of making accurate prediction for small amplitude perturbation and long time running.

\vspace{3ex} {\textbf{Acknowledgments}} \vspace{1ex}
This work was supported by the National Science Foundation of China (11602091, 91530319) and the National Key Research and Development Plan (No. 2016YFB0600805).

\appendix
\renewcommand{\appendixname}{Appendix~}
\section{GKS formula}
\label{app:gks-formula}
Let $g'$ be the normalized equilibrium distribution function,
\begin{eqnarray}
g' &=& \left(\frac{1}{2\pi RT}\right)^{(1+k)/2} e^{-\frac{(u-U)^2+\xi^2}{2RT}}. \label{eq:equilibrium-normalized}
\end{eqnarray}
\subsection{Moments of the Maxwellian distribution function}
\begin{eqnarray*}
\langle u^n\xi^m g \rangle &=& \rho \langle u^n g' \rangle \langle \xi^m g' \rangle
\end{eqnarray*}

\begin{eqnarray*}
\langle u^{n+2} g' \rangle &=& U\langle u^{n+1}g' \rangle + (n+1)RT\langle u^n g' \rangle \\
\langle u^{0} g' \rangle &=& 1 \\
\langle u^{1} g' \rangle &=& U \\
\langle u^{2} g' \rangle &=& U^2 + RT \\
\langle u^{3} g' \rangle &=& U^3 + 3URT \\
\langle u^{4} g' \rangle &=& U^4 + 6U^2 RT + 3(RT)^2 \\
&\cdots\cdots&
\end{eqnarray*}

\begin{eqnarray*}
\langle \xi^0 g' \rangle &=& 1 \\
\langle \xi^1 g' \rangle &=& 0 \\
\langle \xi^2 g' \rangle &=& kRT \\
\langle \xi^3 g' \rangle &=& 0 \\
\langle \xi^4 g' \rangle &=& (k^2+2k)(RT)^2 \\
&\cdots\cdots&
\end{eqnarray*}

\begin{eqnarray*}
\langle -\phi_x g_u \rangle &=& 0 \\
\langle -u\phi_x g_u \rangle &=& \phi_x \rho \\
\langle -u^2\phi_x g_u \rangle &=& 2 \phi_x \rho U \\
\langle -u^3\phi_x g_u \rangle &=& 3 \phi_x \rho (U^2+RT)\\
&\cdots\cdots&
\end{eqnarray*}

\subsection{Derivative of the Maxwellian distribution function}
\begin{eqnarray*}
\frac{\partial g}{\partial s} &=& g \frac{\partial(\ln g)}{\partial s} \\
&=& g \frac{\partial (\ln(\rho) - \frac{1+k}{2}\ln(2\pi RT) - \frac{(u-U)^2+\xi^2}{2RT})}{\partial s},
\end{eqnarray*}
where $s$ represents the space coordinate or time coordinate.
Since $u$ and $\xi$ are independent of coordinate $s$, the derivatives of the Maxwellian distribution function can be expressed by the derivative of macroscopic variables, say, $\frac{\partial \rho}{\partial s}$, $\frac{\partial U}{\partial s}$ and $\frac{\partial T}{\partial s}$, or other set macroscopic variables,
\begin{eqnarray*}
\frac{\partial g}{\partial s} &=& g \left\{ \frac{\partial \rho}{\rho\partial s} - \frac{1+k}{2T}\frac{\partial T}{\partial s} + \frac{(u-U)^2+\xi^2}{2(RT)^2}\frac{\partial T}{\partial s} - \frac{u-U}{RT}\frac{\partial U}{\partial s}\right\}.
\end{eqnarray*}

\subsection{The identities at isothermal hydrostatic equilibrium state}
The isothermal hydrostatic equilibrium state is represented by Eq.(\ref{eq:hydrostaticEq}), at which
the velocity is zero, and $p_x + \rho\phi_x = 0$. Then, we have,
\begin{eqnarray}
\langle \psi(ug_x-\phi_x g_u) \rangle &=&
\left(\begin{array}{c}
(\rho U)_x + 0 \\
(\rho U^2+ p)_x + \rho\phi_x \\
\displaystyle \frac{1}{2}(\rho (U^3+3URT+nURT))_x + \rho U \phi_x
\end{array}\right) = 0, \\
\langle u\psi(ug_x-\phi_x g_u) \rangle &=&
\left(\begin{array}{c}
(\rho U^2+ p)_x + \rho \phi_x \\
(\rho (U^3+3URT))_x + 2\rho U \phi_x \\
\frac{3+k}{2}RT(p_x+\rho\phi_x)
\end{array}\right) = 0.
\end{eqnarray}

\section{Non-conservative discretization}
\label{app:non-conservative}
\begin{figure}[h]
\centering
    \parbox[t]{0.32\textwidth}{
    \includegraphics[totalheight=5cm]{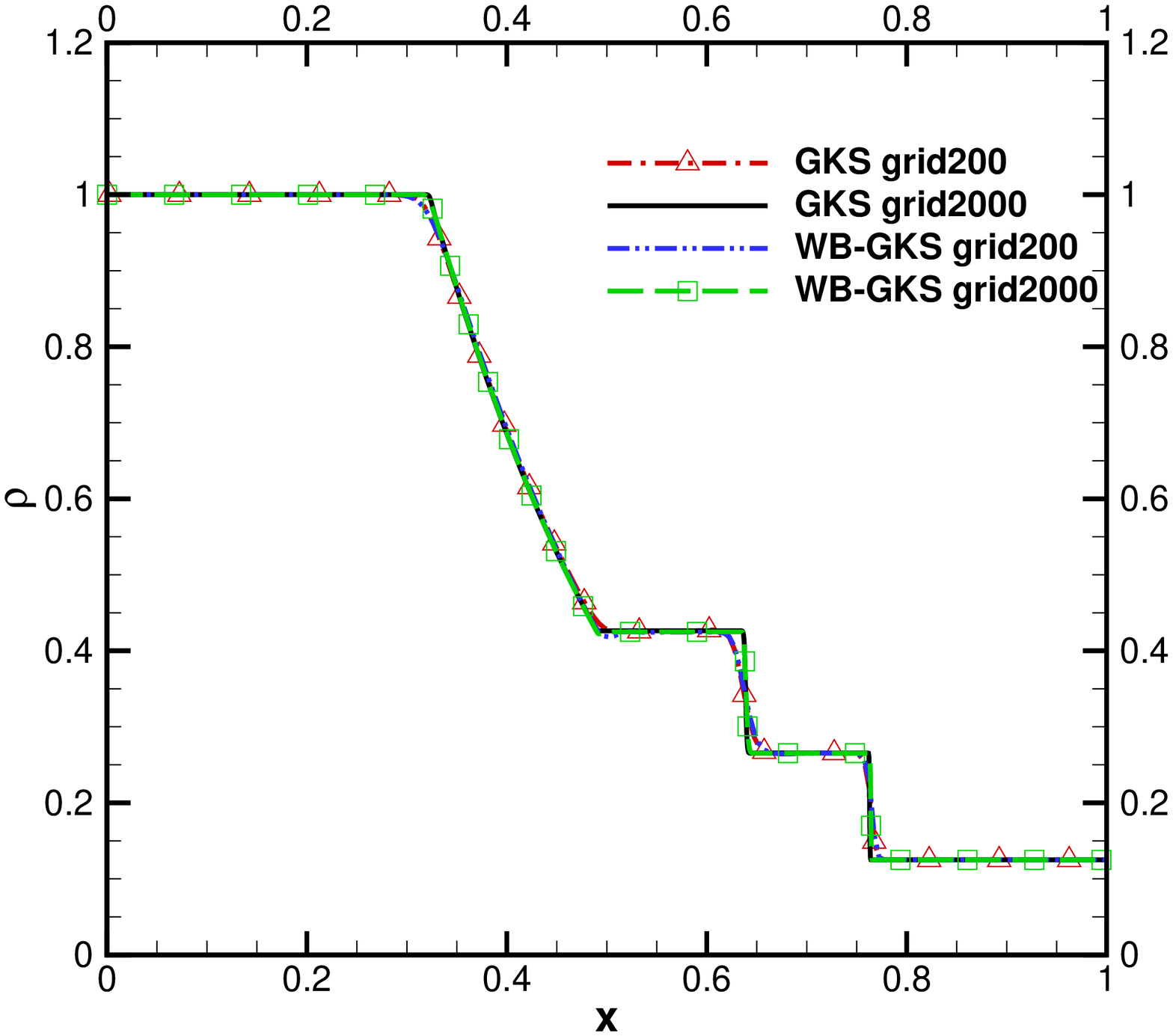}
    }
    \hfill
    \parbox[t]{0.32\textwidth}{
    \includegraphics[totalheight=5cm]{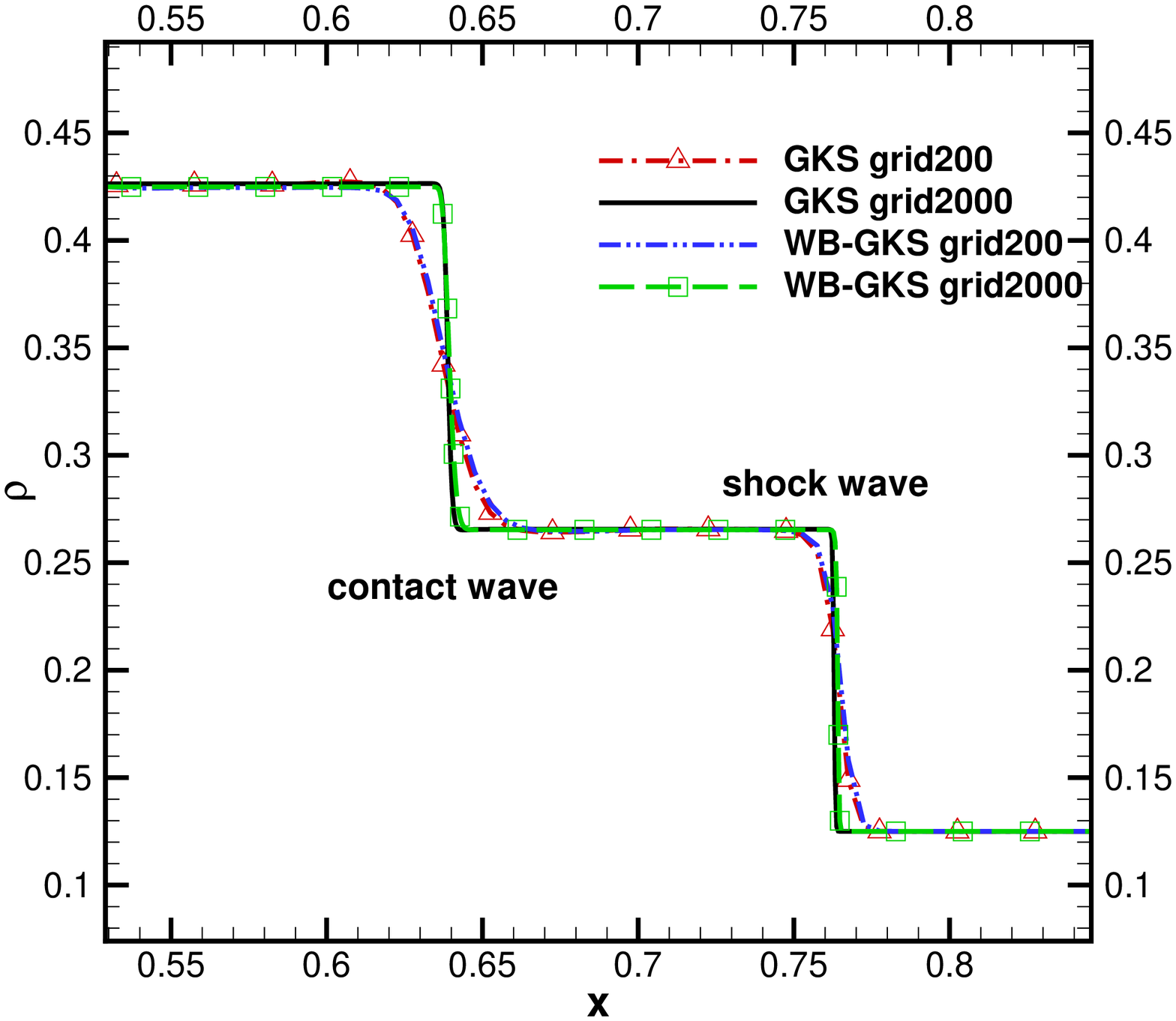}
    }
    \hfill
    \parbox[t]{0.32\textwidth}{
    \includegraphics[totalheight=5cm]{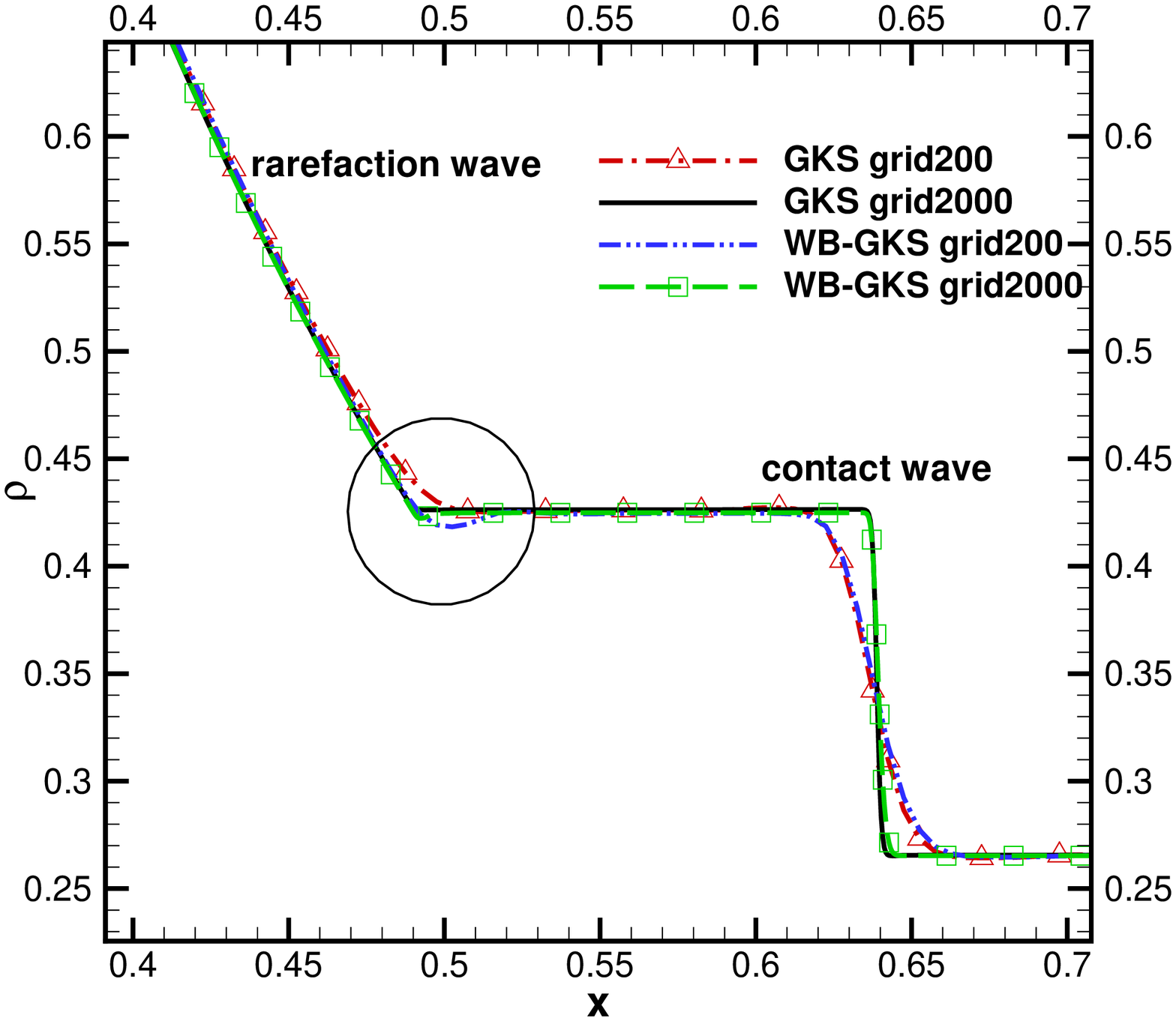}
    }
    \caption{The density distribution of the Sod shock tube at $t=0.15$. The initial condition are $\rho = 1, U = 0, p = 1$ in the interval $[0,0.5]$, and $\rho = 0.125, U = 0, p = 0.1$ in the interval $[0.5,1]$.}
    \label{fig:sod-without-force}
\end{figure}
As aforementioned, a portion of force term is treated as a flux term in the modified momentum equation.
Actually, the proposed scheme is presented in a non-conservative form which may lead to wrong shock speed.
That's why we only consider the smooth flow in this study.
However, the reader might curious about the performance of the proposed scheme on simulating the supersonic flow with shocks.

In order to estimate the effect of our non-conservative scheme, we compare the numerical results of the Sod shock tube without external force. Two numerical schemes, the proposed non-conservative scheme (labeled as WB-GKS) and a conserved scheme (labeled as GKS), are employed. The vanLeer slope limiter is adopted for the interpolation of $\alpha, U, T$. The other discretizations keep unchanged. For more details of high speed GKS, please refer to \cite{xu2001,chen2016cartesian}. The initial conditions are given as follows,
\begin{equation}
\begin{array}{llll}
\rho = 1,& U = 0,& p = 1, &\text{if } x \leq 0.5, \\
\rho = 0.125,& U = 0,& p = 0.1, &\text{if } x > 0.5.
\end{array}\label{eq:sod}
\end{equation}
The kinematic viscosity is $0.0$. The numerical results are shown in figure \ref{fig:sod-without-force}. The overall results are good. It can be seen that the shock speed predicted by our non-conservative scheme is almost identical with result predicted by conserved GKS.
However, there is a little defect near the rarefaction wave as shown in figure \ref{fig:sod-without-force}(c).

\begin{figure}
\centering
    \parbox[b]{0.32\textwidth}{
    Density at $t=0.2$
    \includegraphics[totalheight=3.5cm,bb = 0 26 690 525, clip =
    true]{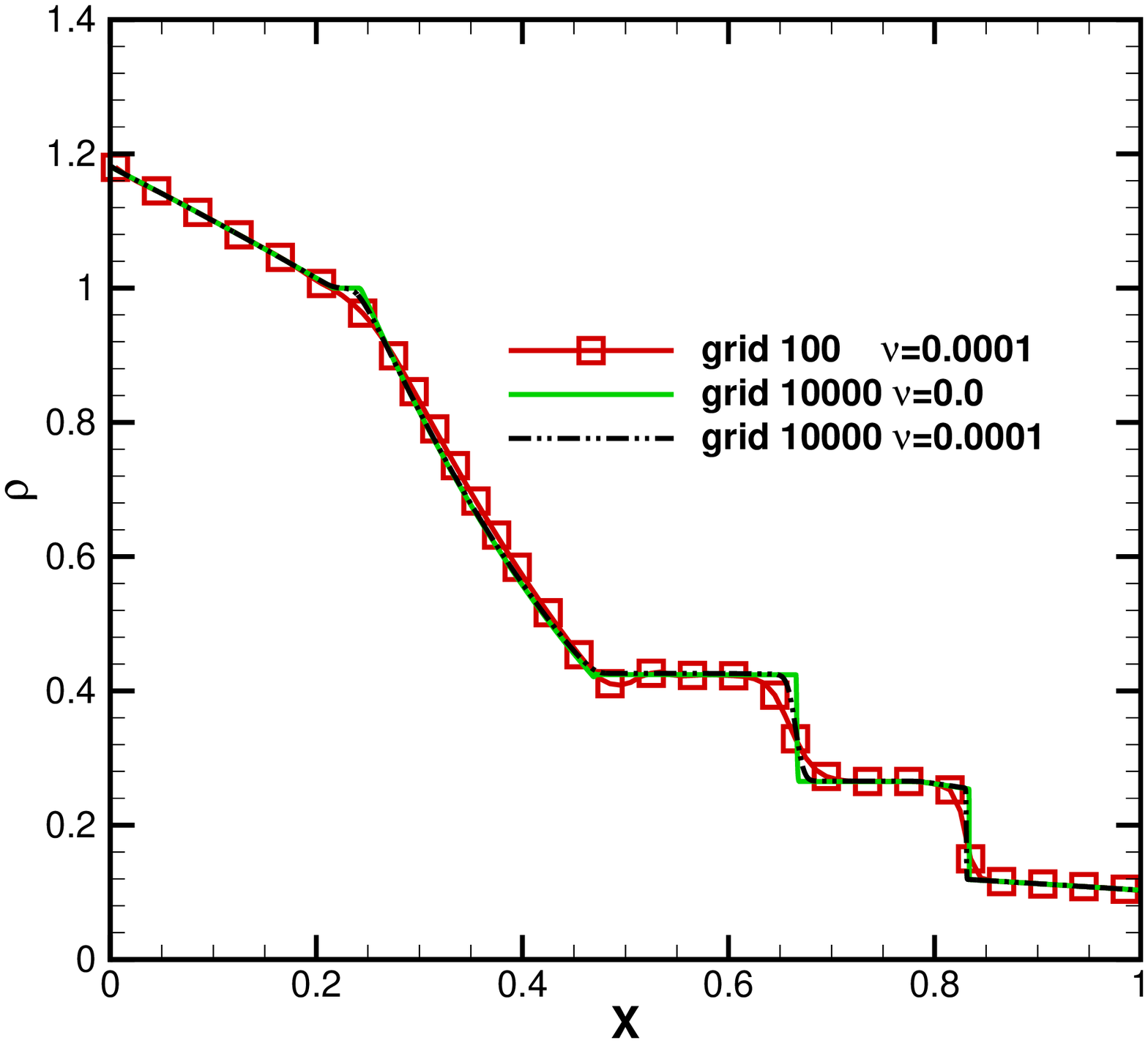}
    }
    \parbox[b]{0.32\textwidth}{
    Pressure at $t=0.2$
    \includegraphics[totalheight=3.5cm,bb = 0 26 690 525, clip =
    true]{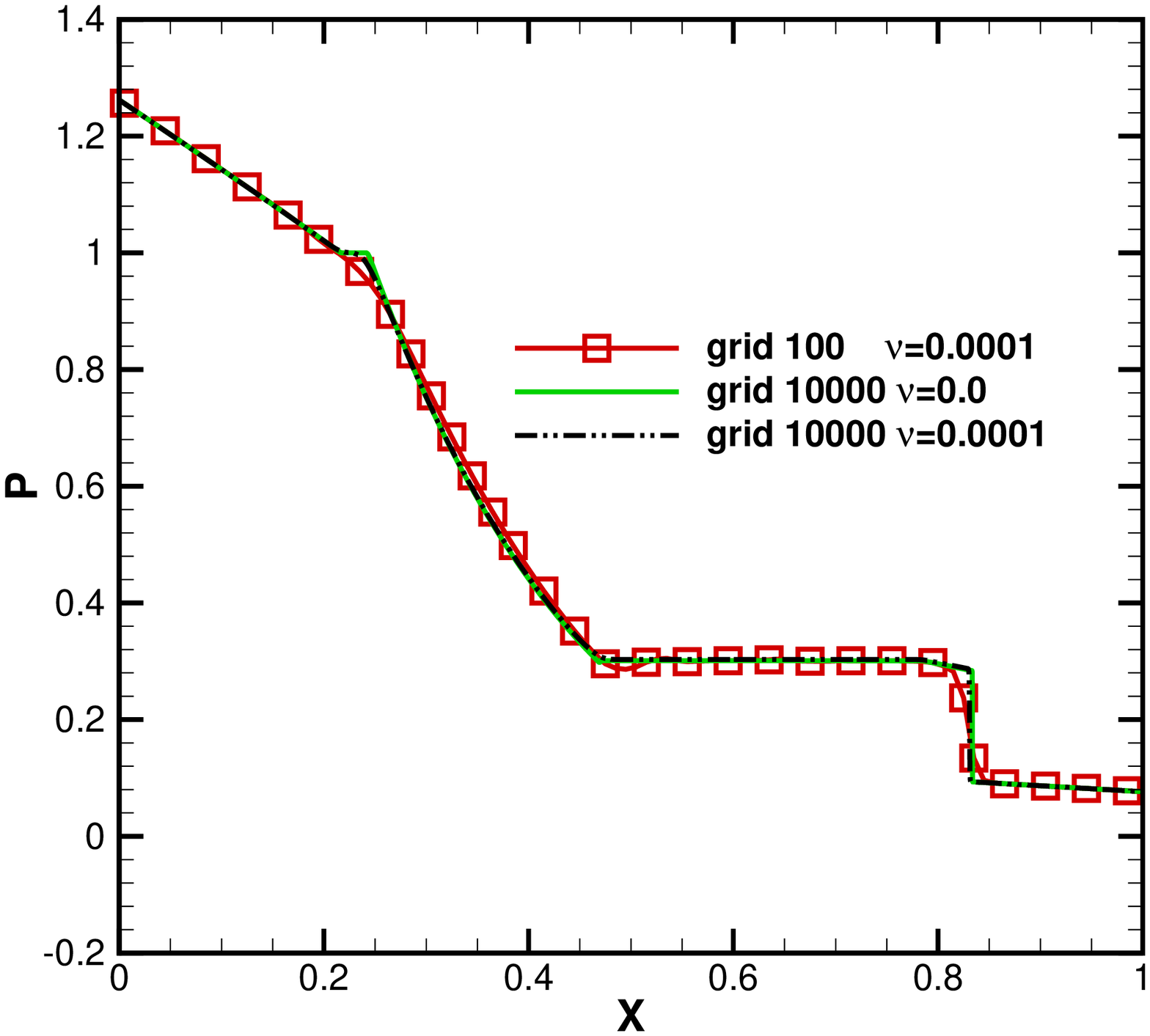}
    }
    \parbox[b]{0.32\textwidth}{
    $t=35000$
    \includegraphics[totalheight=3.5cm,bb = 0 26 690 525, clip =
    true]{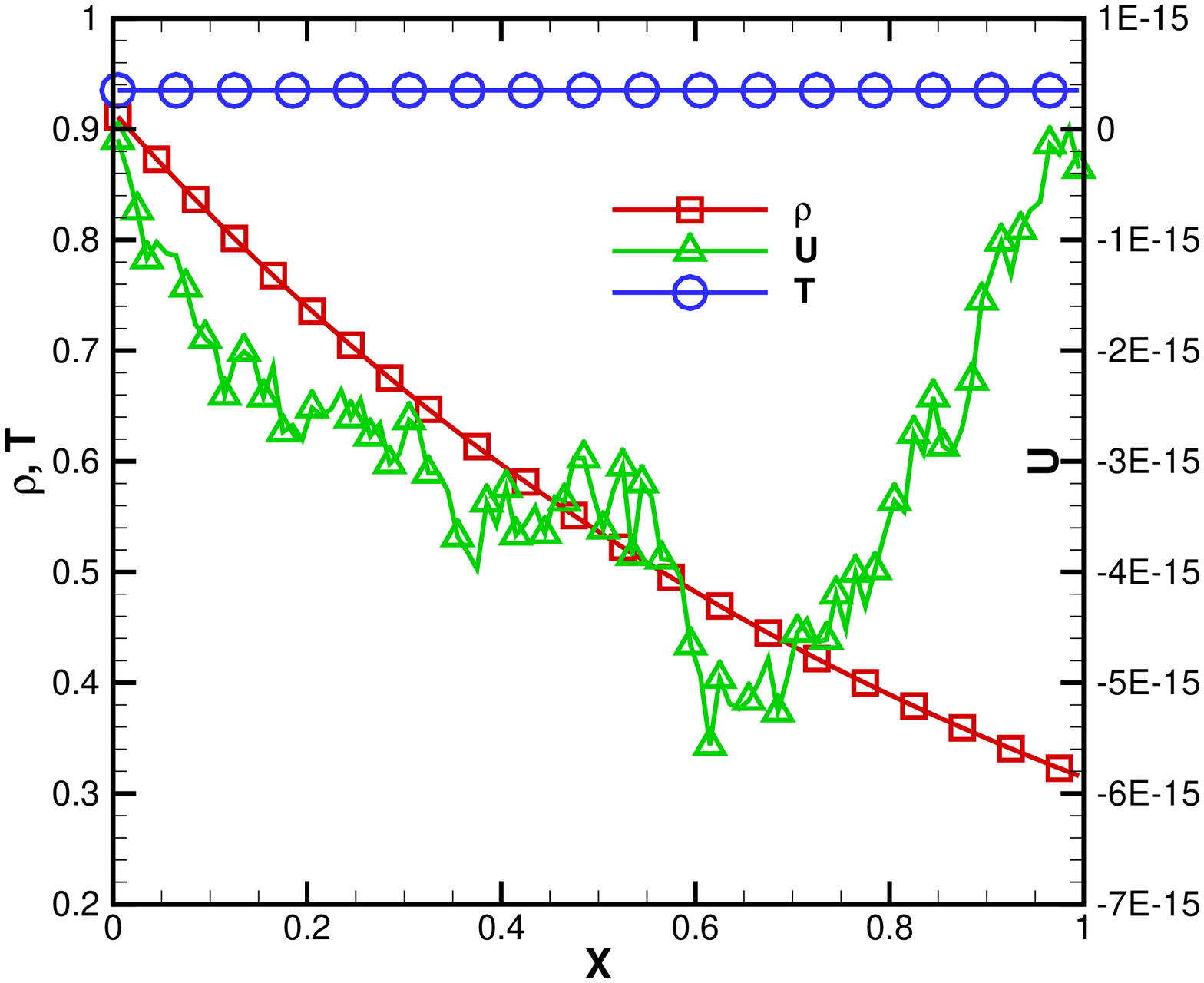}
    }
    \caption{The Sod shock tube problem under linear potential function ($\phi=x+1$). The gases finally settle down to an isothermal hydrostatic equilibrium state with kinematic viscosity $\nu = 0.0001$.}
    \label{fig:sod}
\end{figure}
Then a linear gravitational potential, $\phi = x+1$, is applied.
The kinematic viscosity is $0.0$ and $0.0001$ for the Euler equations and the NS Equations respectively.
The results at $t=0.2$ and $t=35000$ are presented in figure \ref{fig:sod}.
The system eventually return to a quiescent isothermal equilibrium state as expected.

Although, Sod shock tube problem is simulated here, we strongly recommend limiting the application of the proposed scheme to low-speed continuous flow.

\end{document}